\documentclass[letter,12pt]{article}
\usepackage{amssymb}
\usepackage{amsfonts}
\usepackage{graphicx}
\usepackage{amsmath}
\usepackage[authoryear]{natbib}
\usepackage{setspace}
\usepackage{fancybox}
\usepackage[bottom]{footmisc}
\usepackage[left=2cm,top=2.2cm,right=2cm,bottom=2.2cm]{geometry}
\usepackage[normal]{caption}
\usepackage{rotating}
\usepackage{color}
\usepackage{url}
\usepackage{rotating}

\newtheorem{Def}{Definition}

\begin{document}

\title{Modeling the International-Trade Network:\\A Gravity Approach}

\author{Marco Due\~nas\thanks{Institute of Economics, Sant'Anna School of Advanced
Studies, Pisa, Italy. Mail address: Sant'Anna School of Advanced
Studies, Pisa, Italy.Email: \texttt{m.duenasesterling@sssup.it}} \and Giorgio Fagiolo\thanks{Corresponding Author. Institute of Economics, Sant'Anna School of Advanced
Studies, Pisa, Italy. Mail address: Sant'Anna School of Advanced
Studies, Piazza Martiri della Libert\`{a} 33, I-56127 Pisa, Italy.
Tel: +39-050-883282. Fax: +39-050-883344. Email:
\texttt{giorgio.fagiolo@sssup.it}}}

\bigskip

\date{December 2011}

\maketitle

\begin{abstract}
\noindent This paper investigates whether the gravity model (GM) can explain the statistical properties of the International Trade Network (ITN). We fit data on international-trade flows with a GM specification using alternative fitting techniques and we employ GM estimates to build a weighted predicted ITN, whose topological properties are compared to observed ones. Furthermore, we propose an estimation strategy to predict the binary ITN with a GM. We find that the GM successfully replicates the weighted-network structure of the ITN, only if one fixes its binary architecture equal to the observed one. Conversely, the GM performs very badly when asked to predict the presence of a link, or the level of the trade flow it carries, whenever the binary structure must be simultaneously estimated. 

\bigskip


\vskip 0.6cm


\bigskip

\newpage
\end{abstract}

\onehalfspacing

\section{Introduction\label{Section:Introduction}}
The International Trade Network (ITN), aka World-Trade Web (WTW) or World Trade Network (WTN), is defined as the graph representing in each year the web of bilateral-trade relationships between countries in the World. The statistical properties of the ITN, and their evolution over time, have been recently received a lot of attention in a number of contributions.\footnote{See for example \citet{LiC03,SeBo03,Garla2004,Garla2005,Garla2007,serrc07,Bhatta2007b,Bhatta2007a,Fagiolo2008physa,Fagiolo2009pre,Fagiolo2008acs,Fagiolo2009jee,Fagiolo2010jeic,BariFagiGarla2010,Fagiolo_etal_2011physa,DeBene_Tajoli_2011}.} 

Understanding the topology of the ITN is important for two related reasons. First, trade is one of the most important channels of interaction among countries \citep{HelliwellPadmore1985,Krugman1995,Artis2003,Forbes2002}. The knowledge of macroeconomic phenomena such as economic globalization and internationalization, the
spreading of international crises, and the transmission of economic shocks, may be improved by looking at international-trade patterns in a holistic framework, where indirect as well as direct linkages between countries are explicitly taken into consideration \citep{Fagiolo2010jeic}.\footnote{For example,  \citet{AbeFor2005} show that bilateral trade can only explain a small fraction of the impact that an economic shock originating in a given country can have on another one, which is not among its direct-trade partners. Similarly, \citet{Dees_Saint-Guilhem_2011} report that countries that do not trade very much with the U.S. are largely influenced by its dominance over other trade partners linked with the U.S..} Second, ITN topological properties can help to statistically explain macroeconomics dynamics. For example, \citet{Kali_Reyes_2007_growth} and \citet{Kali_Reyes_2010_contagion} have shown that country position in the trade network has substantial implications for economic growth and a good potential for predicting episodes of financial contagion. Furthermore, \citet{Reyes_Schiavo_Fagiolo_2010_jited} suggest that country centrality in the ITN may help to account for the evolution of international economic integration better than what standard statistics, like openness to trade, do. 

The statistical properties of the ITN, in its undirected/directed or binary/weighted characterizations, have been extensively studied and today we know a great deal about the topological architecture of the web of international-trade flows. For example, \citet{SeBo03} and \citet{Garla2004} show that the binary-directed representation of the ITN exhibits a disassortative pattern: countries with many trade partners (i.e., high node degree) are on average connected with countries with few partners (i.e., low average nearest-neighbor degree). Furthermore, partners of well connected countries are less interconnected than those of poorly connected ones, implying some hierarchical arrangements. Remarkably, \citet{Garla2005} show that this evidence is quite stable over time. This casts some doubts on whether economic integration (globalization) has really increased in the last 30 years. Furthermore, node-degrees appear to be very skewed, implying the coexistence of few countries with many partners and many countries with only a few partners.

These issues are taken up in more detail in a few subsequent studies adopting a weighted-network approach to the study of the ITN. The motivation is that a binary approach, by treating all relationship equally, might dramatically underestimate the impact of trade-linkage heterogeneity. This seems indeed to be the case:
\citet{Fagiolo2008physa,Fagiolo2009pre,Fagiolo2009jee} find that the statistical properties of the ITN viewed as a weighted undirected network crucially differ from those exhibited by its binary counterpart. For example, the strength distribution is highly right-skewed, indicating that a few intense trade connections co-exist with a majority of low-intensity ones. This confirms the results obtained by \citet{Bhatta2007b} and \citet{Bhatta2007a}, who find that the size of the group of countries controlling half of the world's trade has decreased in the last decade. Furthermore, weighted-network analyses show that the ITN architecture has been extremely stable in the 1981-2000 period and highlights some interesting regularities \citep{Fagiolo2009pre}. For example, countries holding many trade partners and/or very intense trade relationships are also the richest and most globally central; they typically trade with many partners, but very intensively with only a few of them, which turn out to be very connected themselves; and form few but intensive-trade clusters (i.e., triangular trade patterns).

Most of existing network literature on the ITN, however, has been focusing on a purely empirical quest for statistical properties, largely neglecting the issue of exploring whether theoretical models are able to explain why the ITN is shaped the way it is.\footnote{See \citet{Bhatta2007a} and \cite{Garla2004} for exceptions. See also \citet{Squartini_etal_2011a_pre,Squartini_etal_2011b_pre} for an alternative approach employing null random models that are able to predict whether observed properties of the ITN are statistically meaningful or simply the result of ``constrained'' randomness.}

This paper is a preliminary attempt to fill this gap. We extend the work in \cite{Fagiolo2010jeic} to ask whether the gravity model (GM) can provide a satisfactory theoretical benchmark able to reproduce the observed architecture of the ITN across time. The GM \citep{GravityBook} aims at explaining international-trade bilateral flows using an equation obtained as the equilibrium prediction of a large family of micro-founded models of trade (more on that in Section \ref{Section:DataMeth}). The term ``gravity'' comes about because the predicted relation between trade flows and explanatory variables is similar to Newton's formula: the magnitude of aggregated trade flows between a pair of countries is proportional to the product of country sizes (e.g. the masses, as proxied by country GDPs) and inversely proportional to their geographic distance (interpreted as proxies of trade-resistance factors, e.g. tariffs). From an econometric perspective, the original model-driven prediction can be augmented with a set of country-specific explanatory variables (e.g., population, area, land-locking effects, etc.), as well as with a set of bilateral variables (i.e., geographical contiguity, common language and religion, colony relation, bilateral trade agreements, etc.). The GM can be fitted to the data using different econometric techniques, ranging from simple ordinary least squares (OLS) applied to the log-linearized equation, to two-stage Poisson estimations, employed to correctly deal with the large number of zero trade flows characterizing the data. Overall, the GM is very successful: independent on the technique employed, it typically achieves a very high goodness of fit, e.g. in terms of R-squared coefficients.

Motivated by the well-known empirical success of the GM, we fit data on bilateral trade flows to build a GM-predicted weighted-directed representation of the ITN, which we then compare to the observed one, constructed using original bilateral-flow data. We employ both a static and a dynamic approach. In the static approach, we assume that a GM holds in each subsequent year and we estimate a series of predicted ITN snapshots. In the dynamic approach, we control for time dummies in the estimation to account for change over time and get a unique predicted ITN from the unbalanced panel of predicted flows. In both cases, we end up with a prediction for the expected bilateral-trade flow occurring between any two countries in a given year, and for the probability that a binary link is in place. We complement this information with standard errors of predicted values, so as to evaluate the precision of GM-based estimated quantities. 

In a nutshell, our results suggest that the GM well predicts weighted ITN properties only when the binary structure is kept fixed, equal to the observed one. Conversely, the performance of the GM is very poor when asked to predict ITN weighted properties together with its binary architecture, or when one employs a GM specification to estimate the presence of a link only.   

The rest of the paper is organized as follows. Section \ref{Section:DataMeth} discusses the gravity model and presents data and related methodologies. Our main results are reported in Section \ref{Section:Results}. Finally, Section \ref{Section:Conclusions} concludes and flags some of the challenges facing ITN modeling in the future.

\section{Data and Methodology\label{Section:DataMeth}}

\subsection{Bilateral Trade-Flow Data}
We use international-trade data taken from \citet{Subramanian_Wei_2003_database}, which contains aggregate bilateral imports reported by the IMF Direction of Trade Statistics, measured in U.S. dollars and deflated by U.S. Consumer Price Index at 1982-83 prices. We focus on seven unbalanced cross-sections for the years 1970 to 2000, with a five-year lag. Let $w_{ij}(t)$ be exports from country $i$ to country $j$ in year $t$ and let $N(t)$ the correspondent number of countries reporting at least a positive flow. 

Table \ref{Tab:Summary_Stats} summarizes some descriptive statistics. The number of participating countries and average per-country trade both increase over time. Entry of new countries in the database may be possibly caused either by the availability of new data or by the actual entry of the country in international-trade markets. New trade links, however, seem to increase more than quadratically with the number of participating countries in the last part of the sample, as testified by the rising density.\footnote{Defined as the ratio between $L(t)$ (existing trade partnerships) and $N(t)\cdot[N(t)-1]$ (all possible trade partnerships).} Note also that the number and percentage of countries making up 50\% of total trade seem to remain stable across the years, hinting to a stable core of top traders. Conversely, the percentage of countries controlling 90\% of total world trade has substantially decreased. The concentration process going on in the ITN, despite globalization and international integration, is confirmed also by the decrease in both the number and percentage of flows making up a certain share of total trade.

Given $w_{ij}(t)$ and $N(t)$, we build weight matrices for the correspondent observed trade networks. More precisely: 

\begin{Def}[Observed Weighted ITN]
The observed weighted International Trade Network in a given year $t$ is represented by a weighted-directed graph, where the nodes are the $N(t)$ countries and link weights are fully characterized by the $N(t) \times N(t)$ asymmetric matrix $W(t)$, with entries $w_{ij}(t)$, i.e. exports from country $i$ to country $j$.
\end{Def}

Similarly, one can define the observed binary ITN, where links represent import-export partnerships, as:

\begin{Def}[Observed Binary ITN]
The observed binary International Trade Network in a given year $t$ is represented by a binary-directed graph, where the nodes are the $N(t)$ countries and binary links are fully characterized by the $N(t) \times N(t)$ asymmetric adjacency matrix $A(t)$, with entries $a_{ij}(t)=1$ if and only if $w_{ij}(t)>0$, i.e. exports from country $i$ to country $j$ are strictly positive.
\end{Def}

This database has been studied from a binary/weighted network perspective in \citet{DeBene_Tajoli_2011}. They show that international integration in trade has been increasing over time, but it is still far from being fully accomplished. Indeed, a strong heterogeneity in the profiles of across-country trade partnerships does emerge. This has important implications for both the role of regional trade agreements (i.e., the WTO) and the interplay between extensive and intensive margins of trade \citep{Felbermayr_Kohler_2006_int_ext_margins}.  

In this paper, we take an alternative approach. We characterize the topological properties of the observed ITN and we compare them to the properties displayed by the gravity-based predicted ITN, which we define in the next sub-sections.

\subsection{Gravity-Model Specifications}
The GM, independently proposed by \citet{Tinbergen1962} and \citet{Poyhonen1963}, is the workhorse model to explain bilateral trade flows among countries as a function of import and export market sizes (i.e., GDP) and trade-resistance factors, proxied by geographical distance. The GM derives its name from the functional form linking trade to size and distance, which resembles the expression for the attraction force between two bodies derived by Isaac Newton in classical mechanics. Thus, in analogy with the physics law, it is expected that trade flows increase with the product of some power of country sizes and decrease with some power of geographical distance.

This empirically-inspired law has been found to be consistent with a
number of theoretical foundations \citep{Anderson1979,Bergstrand1985,Deardorff1998,Anderson2003}. In other words, many possibly-conflicting micro-foundations can generate has their equilibrium outcome some gravity-like relation between trade, market sizes and trade-resistance terms.\footnote{This has led \citet{Deardorff1998} to argue that ``just about any plausible model of trade would yield something very like the gravity equation''. See also \citet{Evenett_Keller_2002}.} For example, a gravity-like equation can be derived in trade specialization models, monopolistic-competition frameworks with intra-industry trade, or Hecksher-Ohlin models \citep[see][for comprehensive surveys]{Fratianni2009,benedictis2011trade}.

Notwithstanding the preferred micro-founded explanation, modern empirical interpretations of the gravity expression generalize the original idea including in the formulation a list of additional explanatory variables, covering aspects related to geography, culture, bilateral trade agreements, among others. In Table \ref{table:vars} we report the list of explanatory variables that, following existing literature \citep[see, e.g.,][]{GlickRose2001,RoseSpiegel2002}, we employ in our exercises. GM explanatory variables can be typically grouped in country- or link-specific ones. The former include, in addition to GDP, other country-size proxies like population and geographical area, as well as geographically-related aspects controlling for land-locking effects and continent membership. The latter instead include relational variables characterizing bilateral relationships, as geographical contiguity, colonial ties, regional trade agreements, commonalities in language, colonial history, religion, and currency. Together, these factors have been shown to successfully explain, in a way or in the other, international-trade flows in gravity-equation econometric exercises \citep{GravityBook}.

The most general GM specification that we employ in what follows then reads:
\begin{eqnarray}\label{eq:gravity}
w_{ij}(t) = \alpha_0 Y_i(t)^{\alpha_1} Y_j(t)^{\alpha_2} d_{ij}^{\alpha_3} 
\left[ \prod_{k=1}^K X_{ik}(t)^{\beta_{1k}} X_{jk}(t)^{\beta_{2k}} \right]\times
\\
\times \exp{ \left( 
\sum_{h=1}^H \theta_{h}D_{ijh}(t) + \sum_{l=1}^L (\delta_{1l}Z_{il}+\delta_{2l}Z_{jl}) \right) } \eta_{ij}(t), \nonumber
\end{eqnarray}
where $t$ is the year ($t=1950,1955,\dots,2000$); $w_{ij}(t)$ are export flows from the observed weighted ITN; $i,j=1,...,N(t)$, $i \neq j$; $Y_h(t)$ is year-$t$  GDP of country $h=i,j$ ($i$=exporter; $j$=importer); $d_{ij}$ is geographical distance; $X_{h}(t)$, $h=i,j$, are additional country-size effects (area and population); $D_{ij}$ is a vector of bilateral-relationship variables (contiguity, common language, past and current colonial ties, common religion, common currency, a dummy to control if both countries share a generalized system of preferences, and a regional trade agreement flag); $Z_{i}$ and $Z_{j}$ are country-specific dummies (controlling for land-locking effects and continent membership); finally, $\eta_{ij}(t)$ are the errors (whose mean conditional to explanatory variables obeys $E[\eta_{ij}(t)|\cdot]=1$).

Two remarks are in order. First, note that standard GM specifications assume that any pair of countries trade. In other words, zero trade flows, which are quite frequent in trade data either because of missing values or because the two countries are not trade partners (see Table \ref{Tab:Summary_Stats}), are ruled out from the analysis by the non-linear functional form employed. Therefore, one either log-linearizes Eq. \eqref{eq:gravity} (and excludes observed zero-trade flows) or explicitly deals with over-estimation errors coming from positive GDPs and other country and bilateral variables. In other words, the standard GM specification is not suited to address the issue why any pair of countries that were previously not trading start to trade at some point, or why existing trade relationships terminate. We shall get back to this point in Section \ref{SubSec:Binary}.

Second, and more importantly, we employ a GM specification that slightly differ from \citet{Anderson2003} one, which is one of the most commonly used in GM exercises. \citet{Anderson2003} introduce multilateral resistance terms and importer-exporter fixed effects. Formally, that approach considers that the constant term of the equation \eqref{eq:gravity} must be generalized to a set of importer and exporter dummies. One important implication is that country-size effects are captured by country dummies. This means that characteristics of exporters and importers cannot be generalized \citep{Santos2006}. In any case, all our results are robust to Anderson and van Wincoop's specification. We have therefore chosen to retain the traditional specification because of its more immediate empirical interpretation.

\subsection{Estimation}
Estimation of Eq. \eqref{eq:gravity} is not easy. A straightforward approach consists in log-linearizing the GM specification and apply standard OLS techniques to estimate parameters and obtain predicted values. The existing empirical literature on GM has largely employed this approach \citep[Cf. for example][]{GlickRose2001,RoseSpiegel2002}. 

However, a series of more recent contributions highlighted the risk of biases in estimation induced by OLS applied to log-linear specifications. The main sources of bias come from the treatment of zero-valued flows \citep{Santos2006,Linders2006,Burger2009}, non-linearity and heteroscedasticity \citep{Santos2006}, endogeneity and omitted-term \citep{Baldwin2006}. In particular, the issue of zero-flow treatment is particularly relevant to our analysis. Indeed, log-linearizing the GM equation and applying OLS estimation implies using non-zero trade flows only in the estimation. In network terms, this means that we are keeping the observed binary structure constant (i.e. we are conditioning on adjacency matrices $A(t)$). This is a serious issue if one wants to estimate the presence of a link together with its weight.

To properly account for all these potential difficulties, we can resort to count-data analysis \citep{Long_1997} and fit to the data Poisson pseudo-maximum likelihood models (PMML), either in their standard formulation \citep{Santos2006} or in zero-inflated specifications \citep{Linders2006}. In a nutshell, PPML models allow to estimate Eq. \eqref{eq:gravity} in its original non-linear form, thus avoiding possible correlation between errors and regressors. PPML models use a Poisson distribution to model simultaneously the probability of a zero flow and of a positive (integer) flow. However, it has been noticed that, in the case of international trade, zero flows occur much more frequently than a plain Poisson model would predict \citep{Burger2009}, cf. also Table \ref{Tab:Summary_Stats}. This has led to the family of zero-inflated (ZI) models \citep{Winkelmann_2008_count}. The underlying idea is to model the presence of zeros and positive values as a two-stage process. In this way one treats differently the process of presence-absence of trade partnerships from link-weight determination. In the first stage, one estimates zero-flow probabilities using a standard logit model, and employing a series of regressors that often coincide with those used in the standard GM formulation. In the second stage, conditionally to having non-zero flows, one estimates the magnitude of trade-flow values using either a Poisson (ZIP) or a negative-binomial (ZINB) distribution. Notice that in the second stage there is a non-zero probability of having a zero flow, as the process governing link-weight value determination may attach a zero flow independently on what the first process has done.

To double check our results, we have applied a full range of models to estimate Eq. \eqref{eq:gravity}. In particular, we have employed standard OLS, PPML, ZIP and ZINB approaches. Furthermore, in order to control for dynamic effects, we have estimated
Eq. \ref{eq:gravity} using both a cross-section perspective (i.e., fitting a separate model for each of the 7 waves we end up with in our database) and an unbalanced panel-data approach (i.e., adding time dummies and estimating once and for all the entire data set). We have also controlled for country fixed effects as suggested in \citet{Baldwin2006}. Our results turn out to be very robust to all these alternatives. Therefore in this paper, to avoid redundancy, we report only results from three sets of models (OLS, PPML and ZIP), where a sequence of independent cross sections is estimated without country fixed effects.\footnote{Indeed, ZINB estimates turn out to be very similar to ZIP ones. No dramatic differences are detected between cross-section and panel-data analyses. Similarly, the introduction of country fixed effects do not alter our results below in any crucial ways. Note also that we employ the same set of regressors in both stages of ZIP and ZINB estimates, as listed in Table \ref{table:vars}. Reducing the set of regressors in the first stage does not dramatically change our main results. The whole set of estimation results is available from the authors upon request.} By doing so, we are able to compare a setup where the binary structure of the ITN is kept fixed (OLS) with two alternative setups (PPML and ZIP) where instead one estimates the probability that a link is in place or not, correcting or not for the zero-inflation effect, i.e. when adjacency matrices are endogenous. 

Table \ref{table:gm_est} presents estimation results for year 2000 (similar results hold also for the remaining years) for OLS, PPML and ZIP. Note that, by and large, both signs and orders of magnitude of estimated coefficients do not change with the estimation technique employed. All coefficients have the expected signs, although there are some relevant differences in estimated-coefficient values across methods. For example, OLS estimates for GDP and distance coefficients are quite different from PPML and ZIP ones. The geographical distance coefficient in OLS is negative and stronger than in PPML/ZIP. This may depend on the fact that OLS estimates are computed on a smaller sample (the number of positive observations is almost one half of the whole sample size). OLS differ from PPML/ZIP ones also as far as the importer/exporter GDP elasticity is concerned. In general, GDP elasticities tend to be larger than in other studies as we explicitly consider population and area as additional size effects (entering with a negative sign). This hints to a relevant effect played by per-capita GDP. Furthermore, variables as contiguity, common language, and regional trade agreements enhance trade. In contrast, all variables related to colony relations, common religion and common currency are statistically significant under the OLS models but not so much for PPML.\footnote{Whenever a variable resulted not significant we decided to keep it among the regressors anyway to preserve comparison between estimation techniques.} 

Note also that in (first-stage) logit estimation of the ZIP method, GDP (resp. distance) negatively (resp. positively) affect the probability of having unlinked countries, as expected. Conversely, distance or land-locking effects enhance the probability of missing links. Contiguity coefficient is instead positive: after controlling for geographical distance, sharing a border does not influence the emergence of bilateral trade. This is however a result that does not hold robustly over all cross-sections, where contiguity does not affect significantly the estimated probability. 

Finally, all diagnostic statistics indicate that the estimated models are well-specified \citep{Wooldridge2001} and achieve a quite good (pseudo) $R^2$.

\subsection{The Predicted Weighted ITN}
As long as $Y_i(t)$, $d_{ij}$ and $X_{ik}(t)$ are strictly positive for all $(i,j)$ and $t$, one can rewrite Eq. \eqref{eq:gravity}\footnote{From now on, we suppress time labels for the sake of notational convenience and we refer to a cross-section sequence of estimations.} as:

\begin{equation}
w_{ij}=\exp\{x_{ij}\cdot	\gamma^M\}\eta_{ij},
\end{equation} 
where $x_{ij}$ are logged country-specific and bilateral explanatory variables, and $\gamma$ is the vector of all parameters to estimate. Let $\hat{\gamma}^M$ be estimated parameters with model $M\in\{OLS,PPML,ZIP\}$. 

In the OLS case, therefore, one can straightforwardly define a linear prediction for the log of non-zero flows as:
\begin{equation}
\hat{\omega}_{ij}^{OLS}=\log[\hat{w}_{ij}^{OLS}]=x_{ij}\cdot \hat{\gamma}^{OLS},
\label{OLS_pred}
\end{equation} 
Note that we prefer to use logs of non-zero trade-flows to avoid over-dispersion issues. This means that when comparing observed and OLS-predicted ITN properties we will always refer to logs of non-zero flows to define link-weights. The variance of the prediction ($\hat \sigma_{OLS}^2$) equals the variance of the model (i.e., the sum of squared residuals) divided by the degrees of freedom.

As far as PPML specification is concerned, the probability of observing a certain trade flow is estimated using a Poisson model with expected value equal to the exponential of the linear prediction. Therefore, predicted flows (in levels) read:
\begin{equation}
\hat{w}_{ij}^{PPML}=\exp\{x_{ij}\cdot	\hat{\gamma}^{PPML}\}.
\end{equation} 
Being a Poisson model, the variance of predictions equals their expected value.     

Finally, in the ZIP case, one first estimates the probability $\psi_{ij}$ that a link $(i,j)$ is zero. This is done by fitting a Logit model taking as dependent variable the elements $a_{ij}$ of the adjacency matrices, and as explanatory variables those usually employed to fit a GM equation. Next, for all and only active trade links, one fits $w_{ij}$ with the PPML model above.\footnote{Therefore, the second stage of a ZIP model takes as given the underlying binary structure as in OLS estimation.} The overall probability of observing a given trade-flow level is described by combination of the PPML and the logit processes. As a result, the predicted bilateral flow (in levels) is defined as: 
\begin{equation}
\hat{w}_{ij}^{ZIP}=(1-\hat{\psi}_{ij})\exp\{x_{ij}\cdot \hat{\gamma}^{ZIP}\}=(1-\hat{\psi}_{ij}) \hat \mu_{ij},
\end{equation} 
whereas the variance of the prediction is given by:
\begin{equation}
Var(\hat{w}_{ij}^{ZIP}|x_{ij})=\hat{\psi}_{ij}(1-\hat{\psi}_{ij})[1+\hat{\psi}_{ij}\cdot \hat \mu_{ij}]. \label{eq:var_zip}
\end{equation} 
As a by-product, one can also estimate the overall probability of a zero flow (i.e. of an absent link), which reads $\hat{\psi}_{ij}+(1-\hat{\psi}_{ij})\hat \mu_{ij}$.

Given any of the foregoing predictions for bilateral-trade flows, we then define:   

\begin{Def}[Predicted Weighted ITN]
The predicted weighted International Trade Network, for each given cross-section $t$, is represented by a weighted-directed graph, where the nodes are countries and link weights are fully characterized by the asymmetric matrix $\hat{W}^M$, with entries $\hat{w}_{ij}^M$ and $M\in\{OLS,PPML,ZIP\}$.\footnote{As mentioned, in the OLS case we shall compute weighted network statistics on the logged predicted matrix, whose generic element is $\hat{\omega}_{ij}^{OLS}$.}
\end{Def}

As far as the binary predicted ITN is concerned, two remarks are in order. First, since in the OLS case, as already mentioned, the predicted binary ITN coincides with the observed one, there is no need to address any binary analysis at all. Second, note that by construction estimated trade flows from both PPML and ZIP models are always strictly positive (although in some cases very small). This means that one always ends up with an estimated full binary ITN, which impairs any statistical comparison with the observed binary ITN. Section \ref{SubSec:Binary} discusses these points in more detail.

\subsection{Network Statistics and Confidence Intervals}
We study the extent to which the architecture of the observed ITN over time can be explained by the GM employing a set of standard topological properties (i.e., network statistics), see \citet{Fagiolo2009pre} for a discussion. As Table \ref{table:net_stats} shows, we focus on three families of properties. First, total node-degree and total node-strength, measure, for binary and weighted networks respectively, the number of node partners and total trade intensity. In a directed network, one can also distinguish between node in-degree/in-strength (i.e., number of markets a country imports from, and total imports) and node out-degree/out-strength (i.e., number of markets a country exports to, and total exports). 

Second, total average nearest-neighbor degree (ANND) and strength (ANNS) compute, respectively, the average number of trade partners and total trade value of trade partners of a given node. This gives us an idea of how much a country is connected with other very well-connected countries. ANND and ANNS statistics can be disaggregated so as to account for both import/export partnerships of a country, and import/export partnerships of its partners. More precisely, one can compute four different measures of average nearest-neighbor degree/strength, obtained by coupling the two ways in which a node A can be a partner of a given target country B (importer or exporter) and the two ways in which the partners of A may be related to it (as exporters or importers). Finally, we consider clustering coefficients (CCs), see \citet{Fagiolo2007pre} for a discussion. In the binary case, a node overall CC returns the likelihood that any two trade partners of that node are themselves partners. In the weighted case, these likelihoods are computed taking into account link weights to proxy how strong are the edges of the triangles that are formed in the neighborhood of a node. Again, in the directed case one can disaggregate total node CC according to the four different shapes that directed triangular motifs can exhibit.\footnote{These are labelled \textit{cycle} (if $i$ exports to $j$, who exports to $h$, who exports to $i$), \textit{in} (if both $j$ and $h$, who are trade partners, exports to $i$), \textit{out} (if both $j$ and $h$, who are trade partners, imports from $i$) and \textit{mid} (if $i$ imports from $h$ and exports to $j$, and $j$ and $h$ are trade partners).}

We are interested not only in node average and standard deviation of such statistics over time, but also in the way node statistics correlate, and how such correlation patterns evolve across the years. In particular, we focus on correlation between node degrees (resp., strengths) and ANND (resp., ANNS). This gives us information on the assortativity/disassortativity nature of the ITN. We are also interested in correlation between ND/NS and clustering, to understand the extent to which more and better connected countries trade with partners that trade a lot between them.     

We also derive the variance of predicted network statistics, so as to build their confidence intervals and evaluate the precision of GM estimates. The variance of population averages of a few predicted network statistics can be computed analytically. For instance, given $M\in\{OLS,PPML,ZIP\}$, the predicted population-average of node out-strength ($NS_{OUT}$) reads:
\begin{equation}
\widehat{\overline{NS}}_{out}^{M}=\frac{1}{N}\sum_i\sum_{j\in \hat I_i}\hat w_{ij}^{M},
\end{equation}
where $\hat I_i$ are the predicted export partners of country $i$. The variance of $\widehat{\overline{NS}}_{out}^{M}$ equals the sum of variances of the elements $\hat w_{ij}^{M}$ divided by $N^2$, as by construction the predictions $\hat w_{ij}^{M}$ have zero covariance. Then, for the PPML model we get:
\begin{equation}
Var(\widehat{\overline{NS}}_{out}^{PPML})=\frac{1}{N^2}\sum_i\sum_{j\in \hat I_i}\hat w_{ij}^{PPML}=\frac{1}{N}\widehat{\overline{NS}}_{out}^{PPML},
\end{equation}
as it is expected from the Poisson nature of the model. In the case of a ZIP estimation, using Eq. \eqref{eq:var_zip}, one instead gets: 
\begin{equation}
Var(\widehat{\overline{NS}}_{out}^{ZIP})=\frac{1}{N^2}\sum_i\sum_{j\in \hat I_i}\hat \mu_{ij}(1-\hat\psi_{ij})(1+\hat\mu_{ij}\hat\psi_{ij}).
\end{equation}
In the OLS case the variance of $\hat w_{ij}$ equals $\hat \sigma_{OLS}^2$ for each observation. Hence:
\begin{equation}
Var(\widehat{\overline{NS}}_{out}^{OLS})
= \frac{\hat \sigma_{OLS}^2}{N^2}\sum_{ij} a_{ij} = \frac{\rho\hat\sigma_{ols}^2(N-1)}{N},
\end{equation}
where $\rho$ is the observed density of the network.

More generally, whenever an analytical expression cannot be obtained, we perform simulations to estimate the variance of predicted network-statistics. To that end, for each model $M\in\{OLS,PPML,ZIP\}$, we generate 10,000 replications of the ITN using the data-generation process implied by the model $M$ to proxy the second moment of the distribution of predicted statistics. For example, in the OLS model the generic entry of the sampled ITN weight matrices in each run can be simply drawn from a normal distribution with mean $\hat \omega_{ij}^{OLS}$ and variance $\hat \sigma_{OLS}^2$, see Eq. \eqref{OLS_pred}. In a PPML model, instead, $\tilde w_{ij}$ are drawn from their corresponding Poisson distribution: Prob$\{\tilde w_{ij}\}=\hat\mu_{ij}^{\tilde w_{ij}}e^{-\hat\mu_{ij}}/\tilde w_{ij}!$, where $\hat\mu_{ij}=\exp(x_{ij}\hat\gamma^{PPML})$. Finally, simulation of ZIP-generated trade flows proceeds in two stages. First, we simulate the binary structure using a Bernoulli process, where each link is in place with probability $(1-\hat\psi_{ij})$. Second, we superimpose on this simulated binary structure a weight matrix whose generic entries are sampled from a Poisson distribution with parameter $\hat\mu_{ij}=\exp\{(x_{ij}\hat\gamma^{PPML})\}$.\footnote{This simulation technique allows the Bernoulli-generated binary structure to converge to the correspondent logit predictions from the first stage of the ZIP. In small samples, however, a bias is introduced. This in general implies an overestimation of the performance of the model. Notice also that in each simulated instance the adjacency matrix changes and so do binary topological properties.}

\section{Results\label{Section:Results}}
This Section explores the question whether the statistical properties of the predicted ITN are similar to those observed in the real-world ITN. We start with basic (non-directed) weighted statistics (total NS, ANNS and clustering). Next, we discuss results related to directed weighted measures (e.g., in and out strength, etc.). Finally, we focus on the binary ITN.
  
\subsection{Weighted Statistics}
\begin{figure}[t]
	\begin{center}
	\begin{minipage}[t]{5.6cm}
		\includegraphics[width=5.6cm]{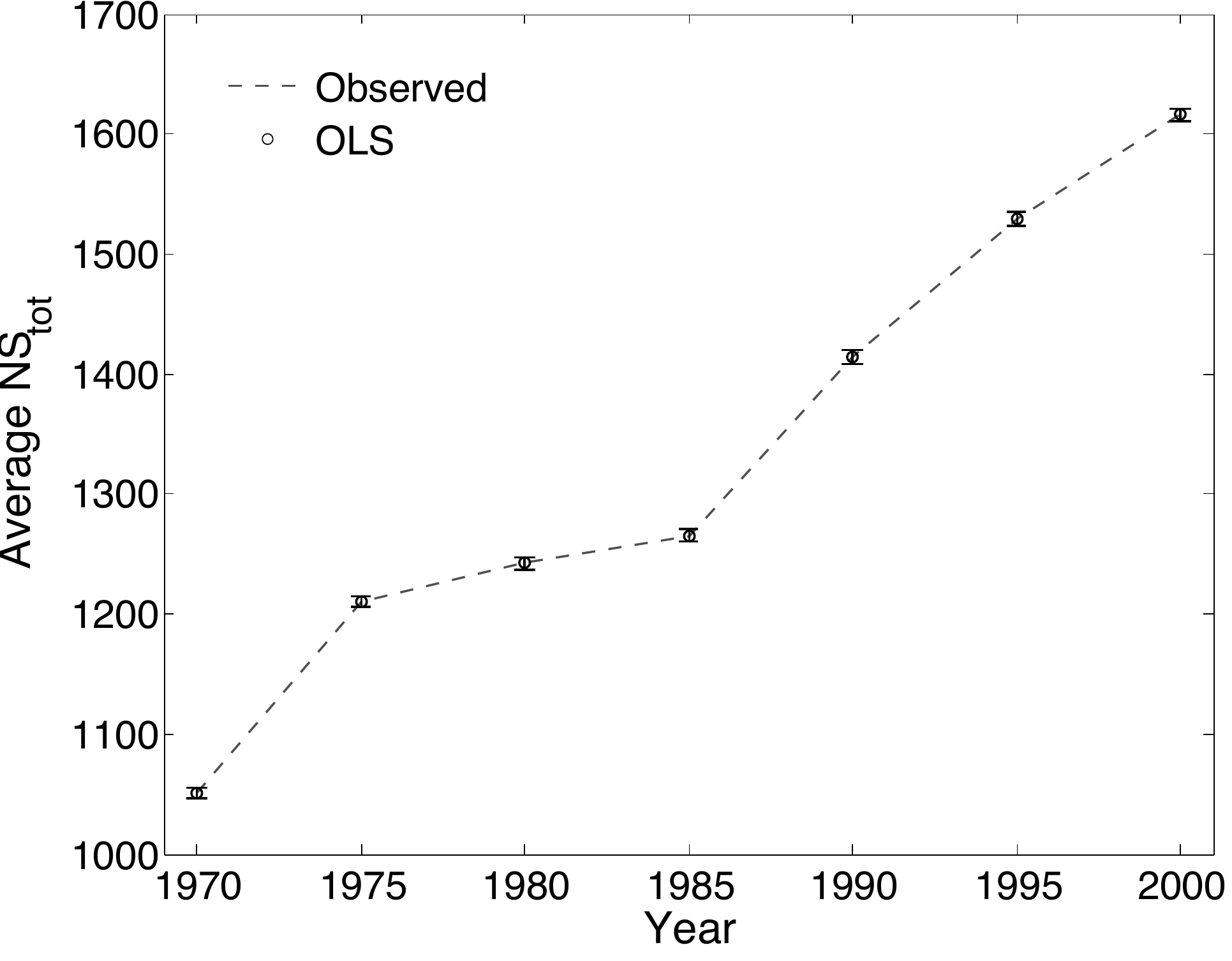}
	\end{minipage}
	\begin{minipage}[t]{5.6cm}
		\includegraphics[width=5.6cm]{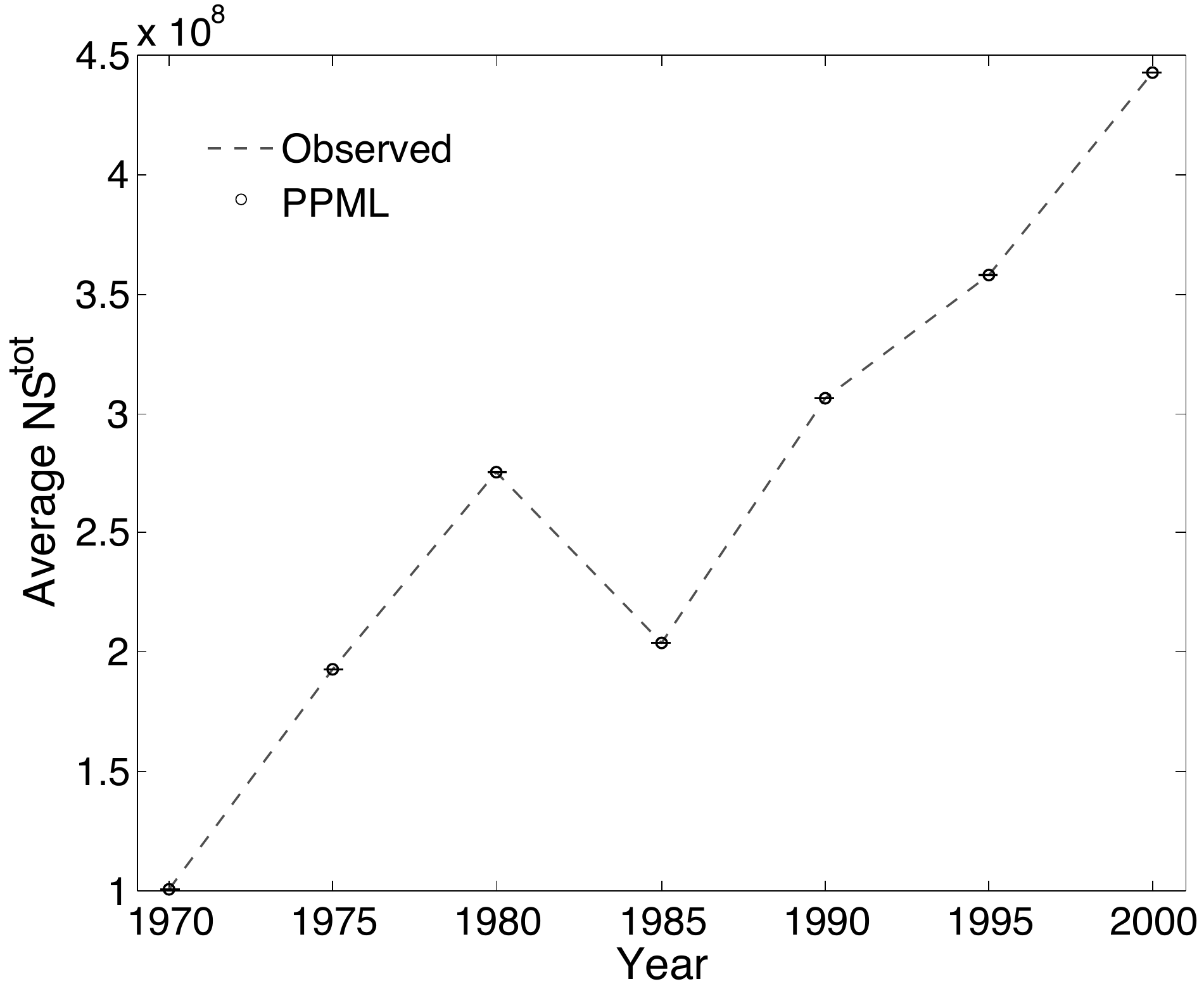}
	\end{minipage}
	\begin{minipage}[t]{5.6cm}
		\includegraphics[width=5.6cm,height=4.6cm]{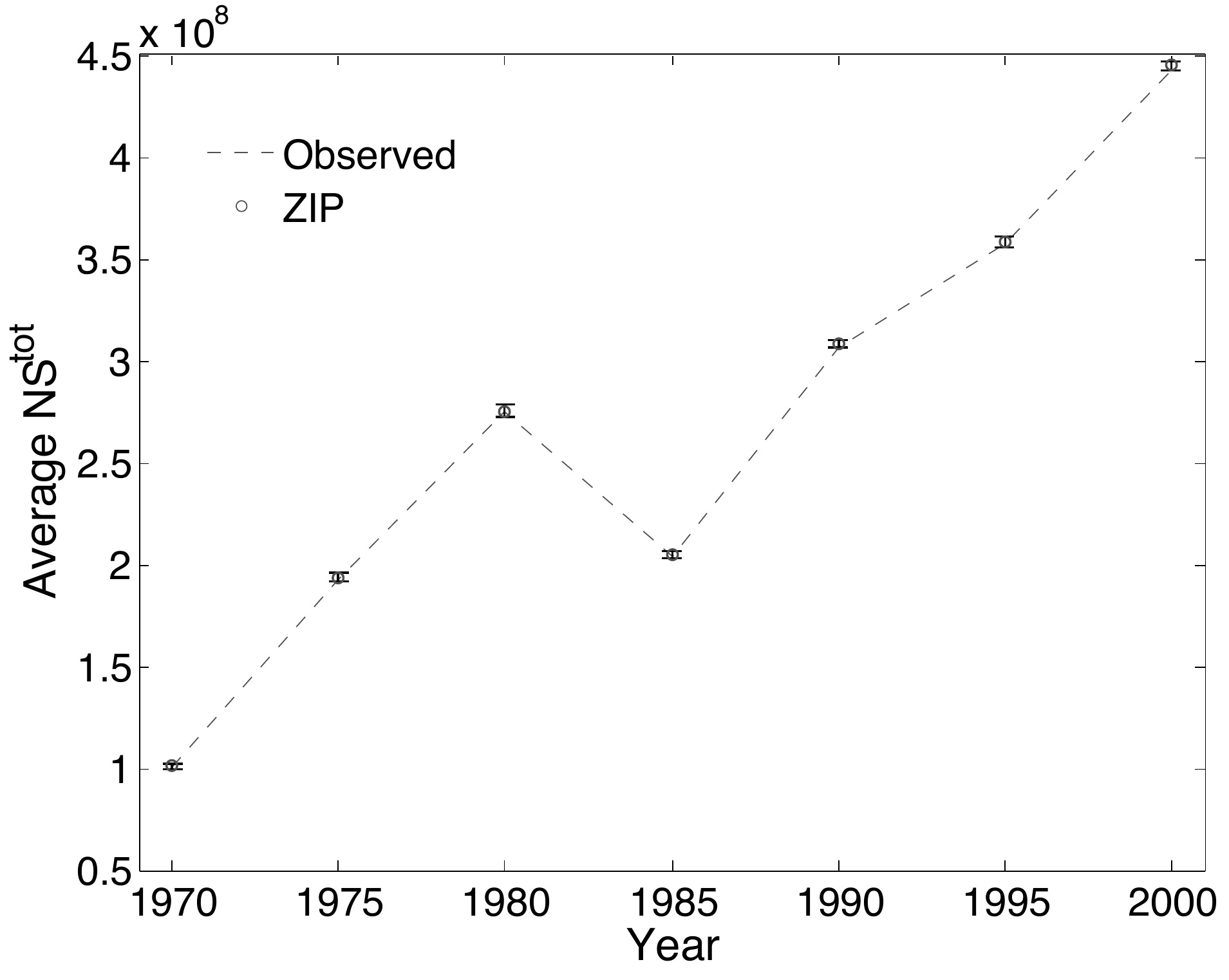}
	\end{minipage}
	\end{center}
	\caption{Observed vs. GM-predicted average node total strength. 95\% confidence bands are displayed as error bars around predicted values. \textit{Note}: in the OLS plot node strength is computed using log of trade flows.} \label{Fig:av_ts}
\end{figure}

We begin to study population averages of total node strength:

\begin{equation}
\widehat{\overline{NS}}_{tot}^{M}=\frac{1}{N}\sum_i{\widehat{NS}_{i,tot}^{M}}=\frac{1}{N}\sum_i\sum_{j}\hat w_{ij}^{M},
\end{equation}
where $N$ is the number of countries in the target cross section. Note that $NS_i^{tot}$ measures total country trade. Therefore its population average equals total world trade divided by the number of countries. Figure \ref{Fig:av_ts} reports predicted values with confidence bands against observed ones across years. It is easy to see that all three methods perfectly match observed values, with very narrow prediction errors. This is not surprising, as the very purpose of the GM is to predict bilateral trade flows, and NS are just linear combinations of them. Therefore one expects the GM to be well equipped to predict linear transformations of total world trade.

\begin{figure}[t]
	\begin{center}
	\begin{minipage}[t]{5.6cm}
		\includegraphics[width=5.6cm]{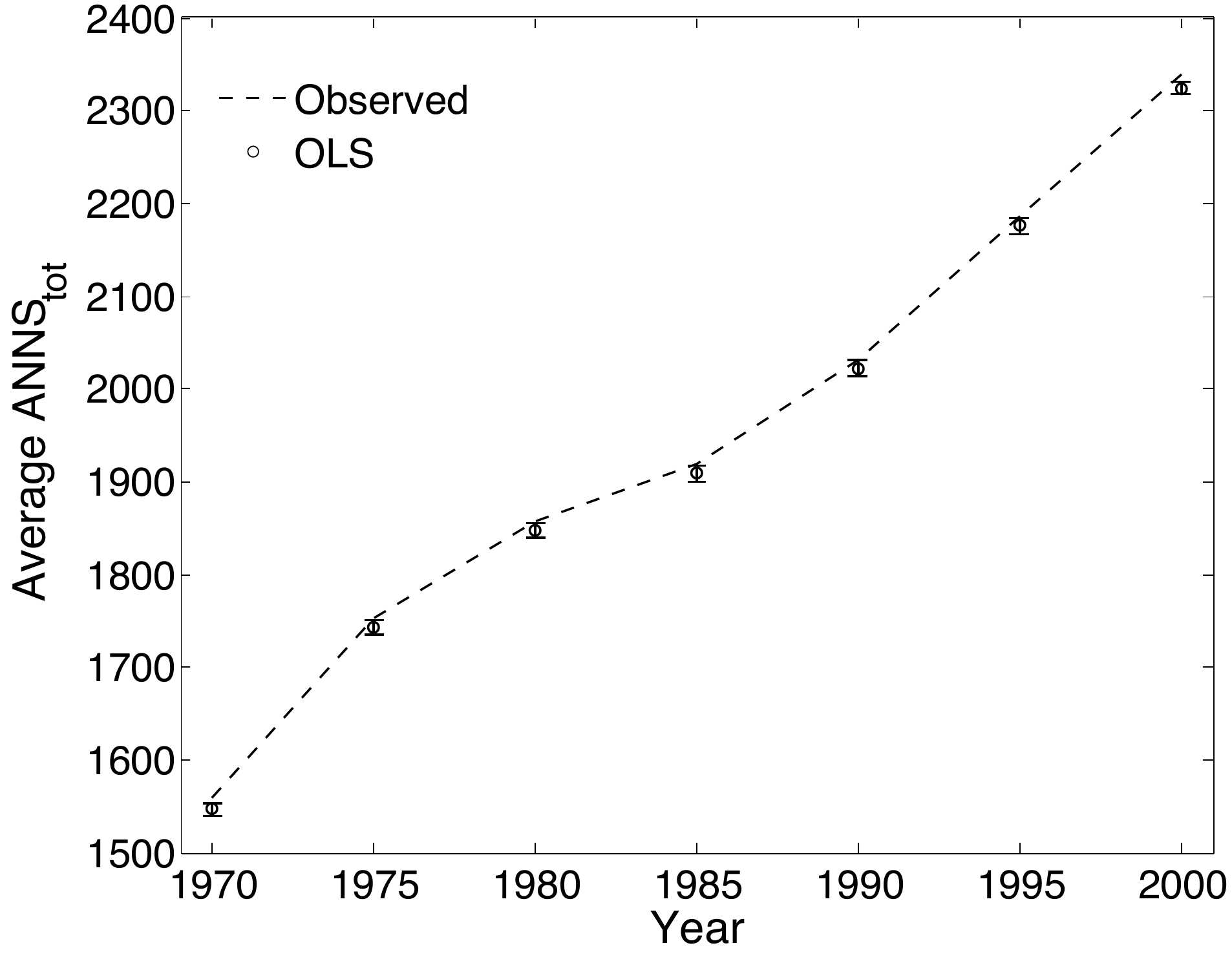}
	\end{minipage}
	\begin{minipage}[t]{5.6cm}
		\includegraphics[width=5.6cm]{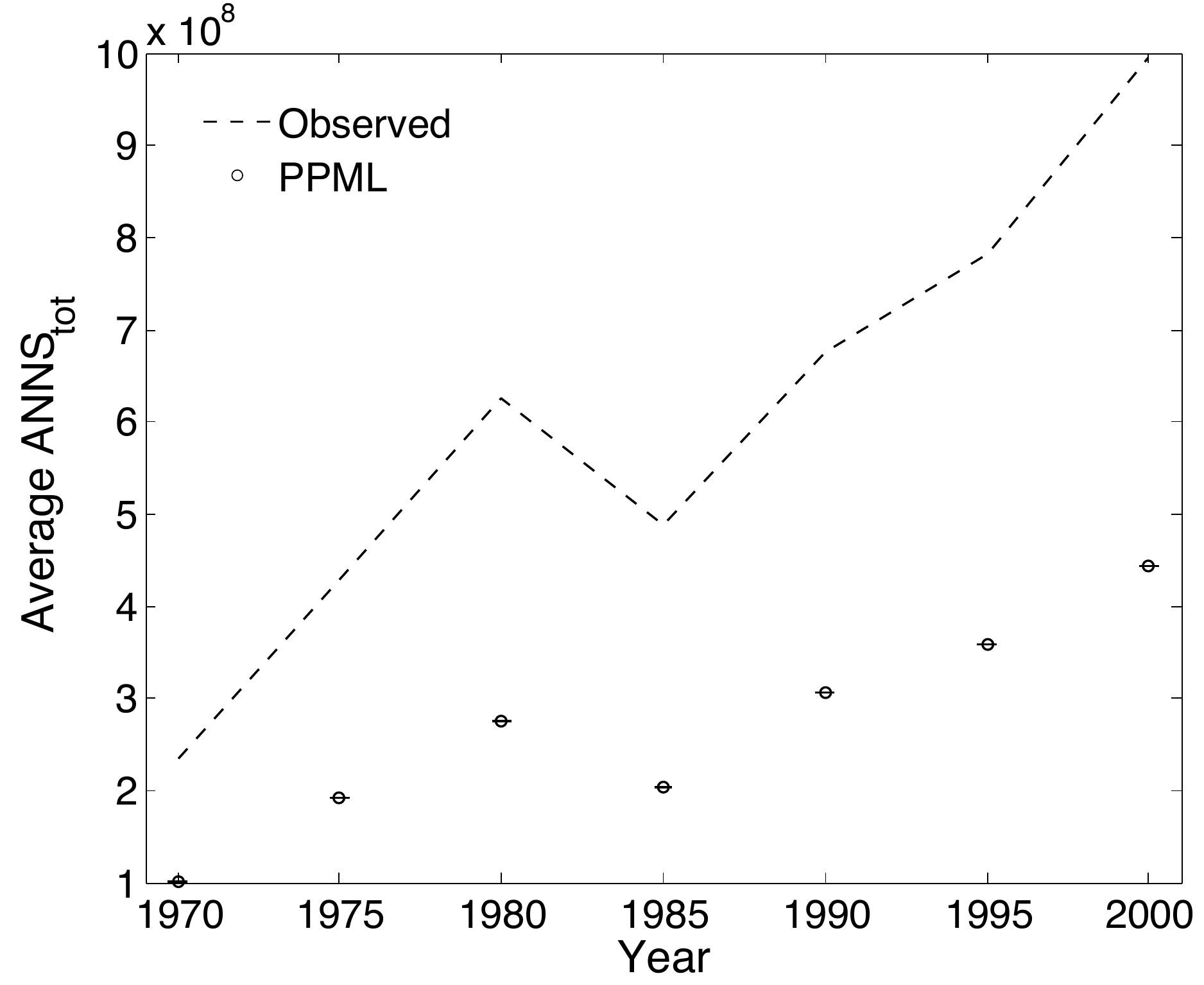}
	\end{minipage}
	\begin{minipage}[t]{5.6cm}
		\includegraphics[width=5.6cm]{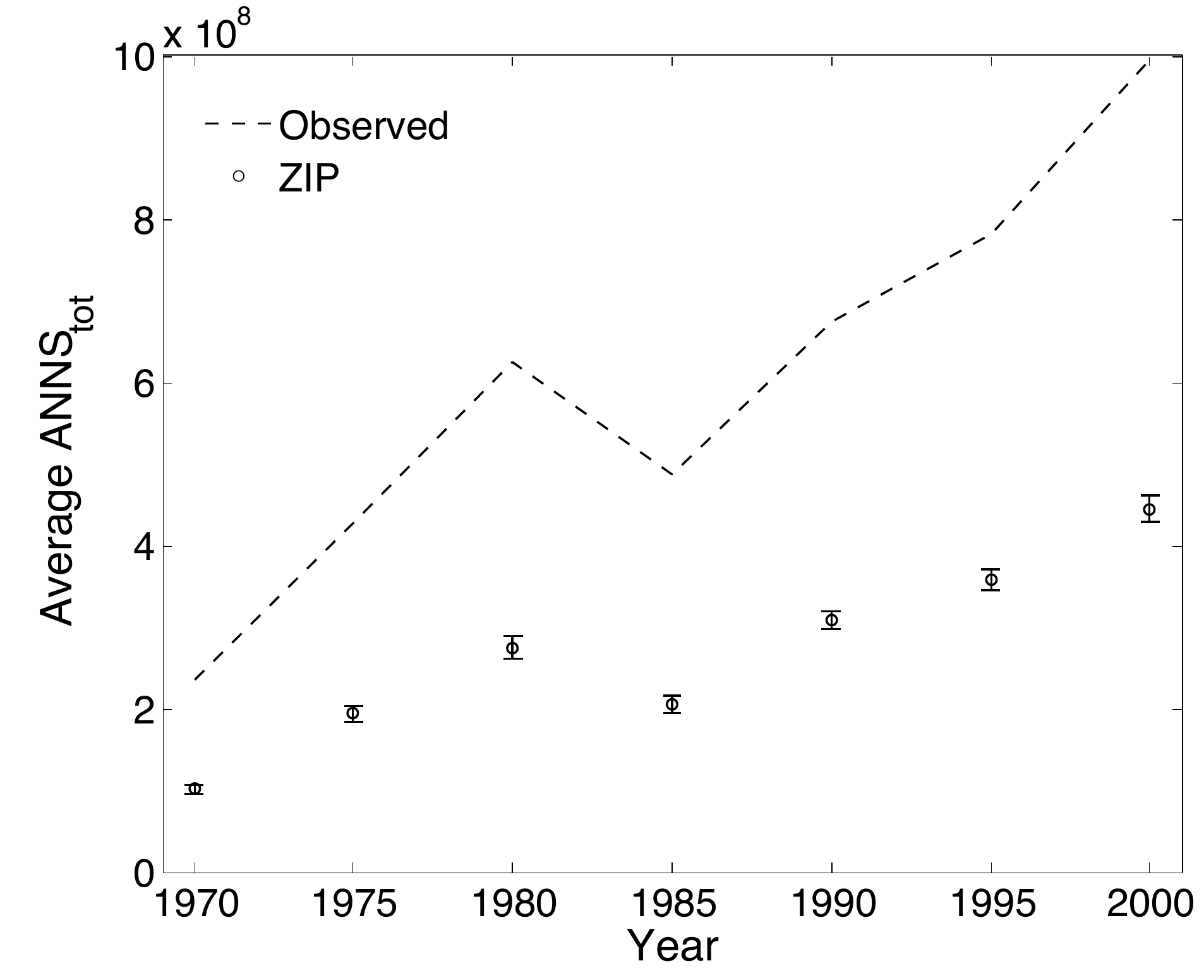}
	\end{minipage}
	\end{center}
	\caption{Observed vs. GM-predicted average total ANNS. 95\% confidence bands are displayed as error bars around predicted values. \textit{Note}: in the OLS plot node strength is computed using log of trade flows.} \label{Fig:av_annst}
\end{figure}

\begin{figure}[t]
	\begin{center}
	\begin{minipage}[t]{5.6cm}
		\includegraphics[width=5.6cm]{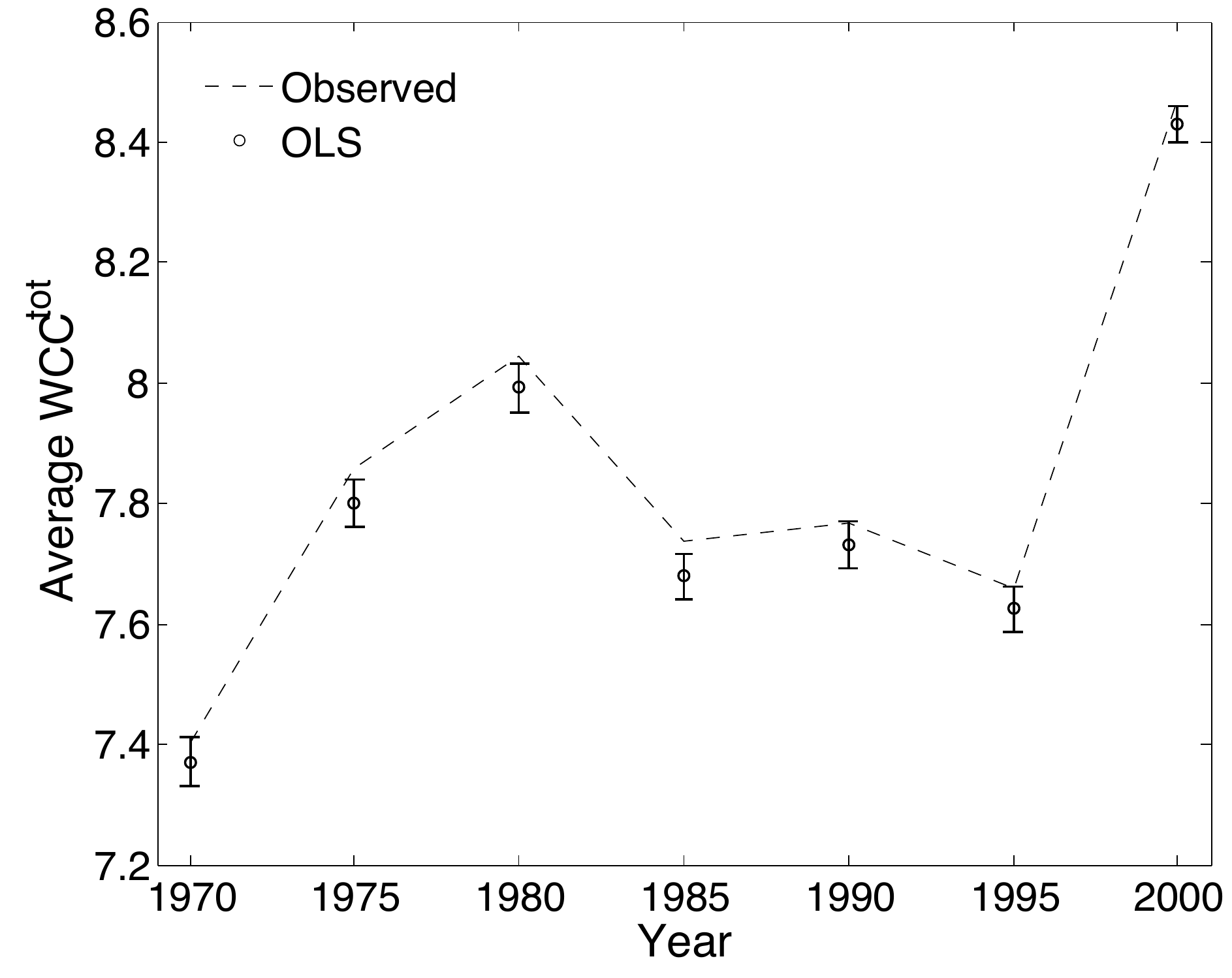}
	\end{minipage}
	\begin{minipage}[t]{5.6cm}
		\includegraphics[width=5.6cm]{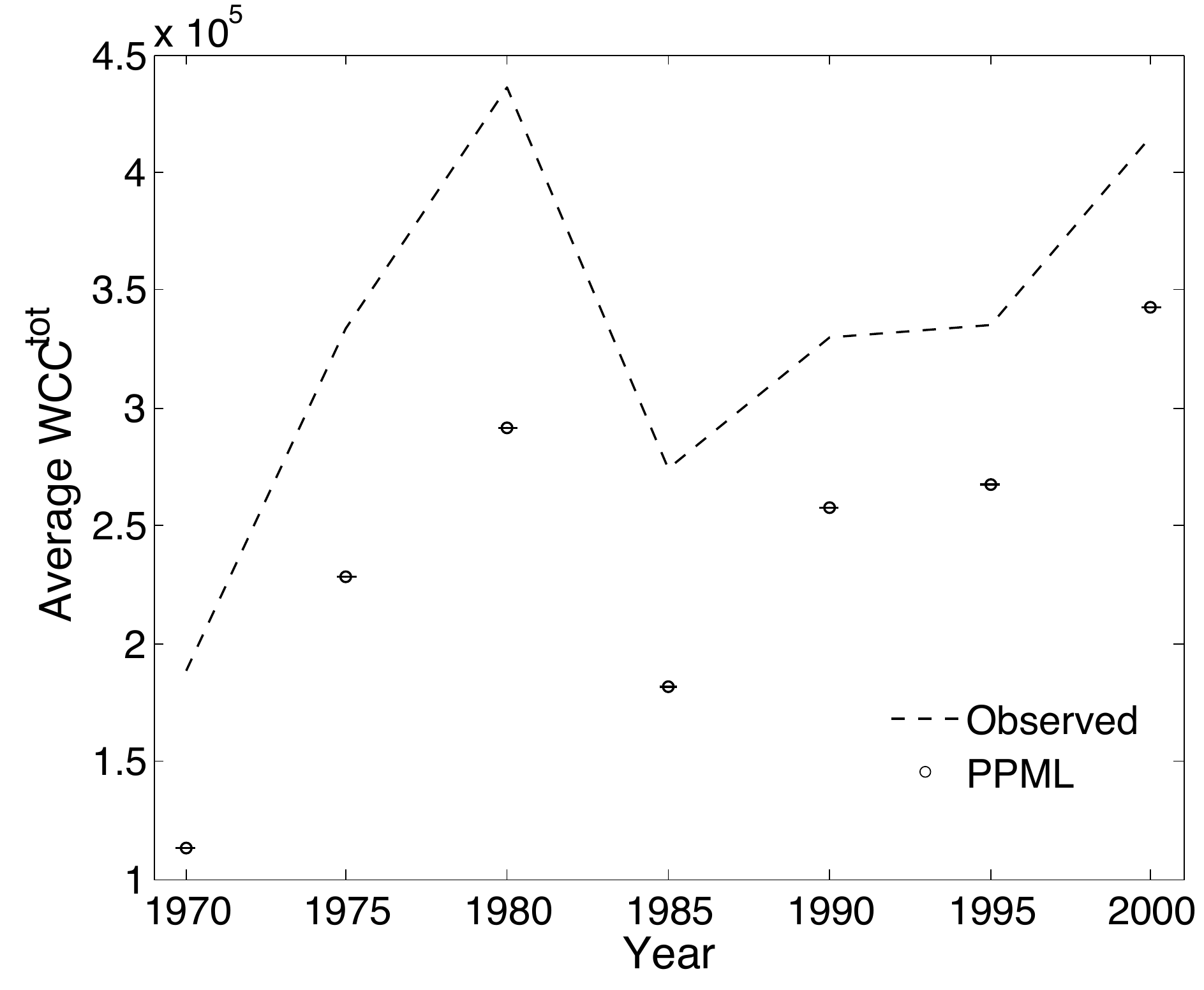}
	\end{minipage}
	\begin{minipage}[t]{5.6cm}
		\includegraphics[width=5.6cm]{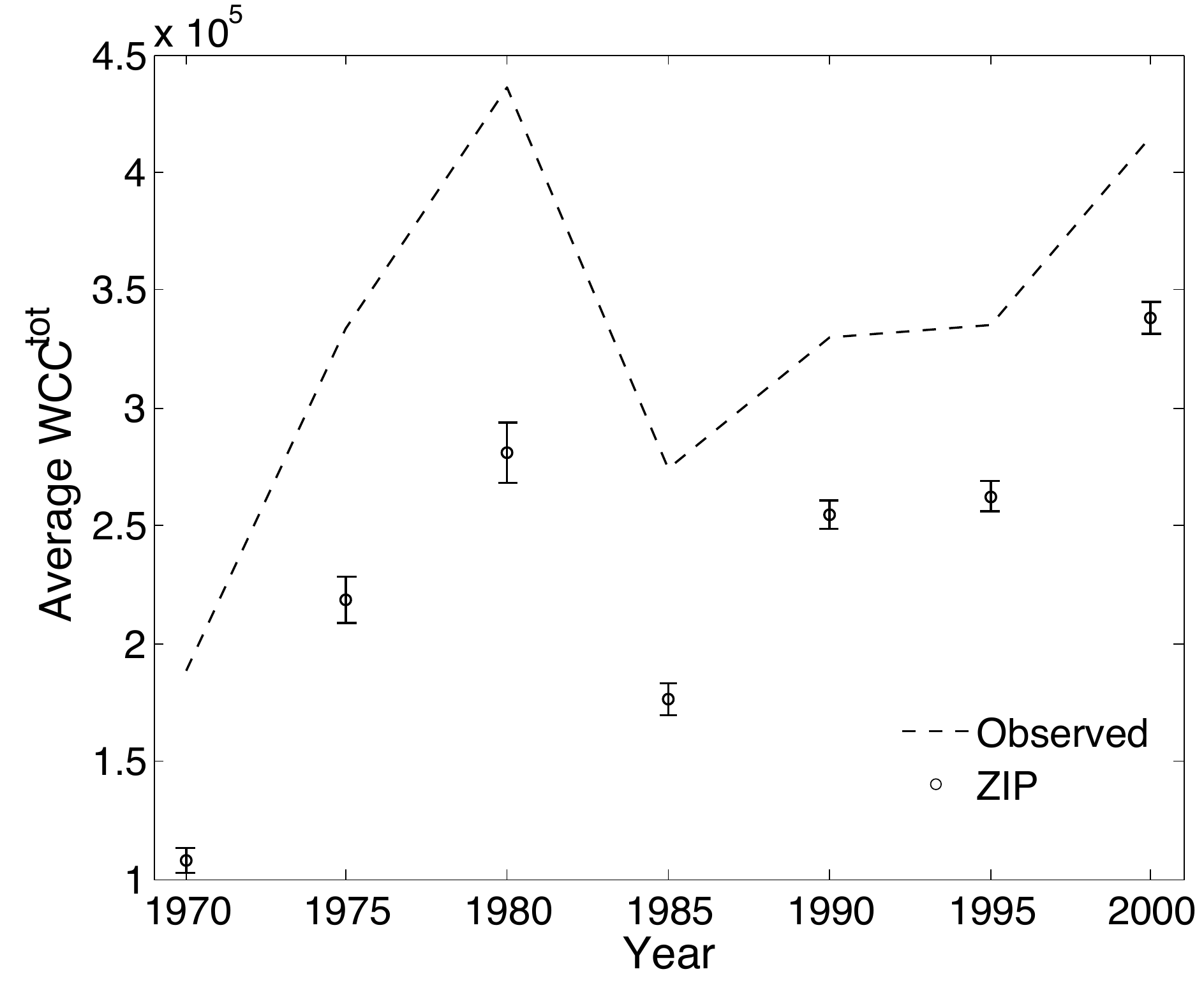}
	\end{minipage}
	\end{center}
	\caption{Observed vs. GM-predicted average total WCC. 95\% confidence bands are displayed as error bars around predicted values. \textit{Note}: in the OLS plot node strength is computed using log of trade flows.} \label{Fig:av_wcct}
\end{figure}

The picture substantially changes when we turn to higher-order statistics like ANNS and WCC, which involve link weights that are two steps away from the origin node. As Figures \ref{Fig:av_annst} and \ref{Fig:av_wcct} indicate, OLS predictions are quite successful in replicating average total ANNS and, to a lesser extent, average total clustering.\footnote{Clustering coefficients are computed without rescaling link weights in the unit interval in order not to bias the analysis with network-dependent rescaling factors \citep{Fagiolo2007pre,kertesz_definitionclustering}. Therefore, the range of WCC is not within $[0,1]$.} More precisely, the OLS-predicted ITN tends to slightly overestimate observed average total ANNS and to underestimate observed average total WCC. Nevertheless, predicted values are very close to (and in many cases within) error bands. Note also that the precision of GM estimates is very high, as the narrow 95\% error bars suggest. Conversely, both PPML and ZIP largely underestimate both ANNS and WCC, although they are able to correctly get the time trend. In addition, PPML predictions are more precise than ZIP ones.

The reason for this mismatch lies in the way the three estimation techniques work in reproducing the binary structure. Recall from Table \ref{table:net_stats} that weighted-network statistics as ANNS and WCC are in fact a mix of link weights and node degrees. To correctly reproduce such network properties any predictor of link weights must also correctly reproduce the underlying binary topology. Using OLS means fitting the GM over positive link weights only, i.e. the observed binary topology is preserved. PPML and ZIP employ instead all possible country pairs in the regression. As a result, they both completely destroy the underlying binary structure and obtain a full predicted binary network, where all links are in place. This is because all predicted bilateral flows and positive-link probabilities are strictly positive (although sometimes very small).\footnote{This interpretation is confirmed by an additional fitting exercise where we employ a PPML estimation technique performed on positive link weights only. In this case, binary-restricted PPML fits are able to reproduce quite successfully all statistics, in line with what happens for OLS. In fact, restricted PPML predictions do not add much more precision in the estimates compared to OLS.} Note that density in ITN ranges from 0.40 to 0.50 (see Table \ref{Tab:Summary_Stats}), meaning that slightly less than a half of possible trade relationships are present. In fact, both PPML and ZIP predict very small link-probabilities for unconnected countries, and consequently very weak link weights. This implies that predicted total country trade (i.e., total node strength) is not dramatically far from the observed one, although in principle it may be slightly overestimated. Conversely, badly predicting the binary structure results in remarkably-smaller predictions for average ANNS and WCC (as compared to observed ones), because one highly overestimates node degrees, appearing at the denominator of both statistics.

To further explore this issue, we perform two-sample Kolmogorov-Smirnov (K-S) tests to compare predicted vs. observed node-statistic distributions. More precisely, given any of the three statistics of interest (total node strength, average nearest-neighbor strength and weighted clustering), we test the null hypothesis that predicted and observed statistics come from the same distribution. The results in Table \ref{table:ks_tests} confirm the message coming from population averages. Consistently over the years, OLS predictions are able to generate network-statistic distributions very similar to those observed in the observed ITN. In contrast, PPML and ZIP can reproduce total strength, but they hardly replicate second-order topology measures like ANNS and WCC.

\begin{figure}[t]
	\begin{center}
	\begin{minipage}[t]{5.7cm}
		\includegraphics[width=5.7cm]{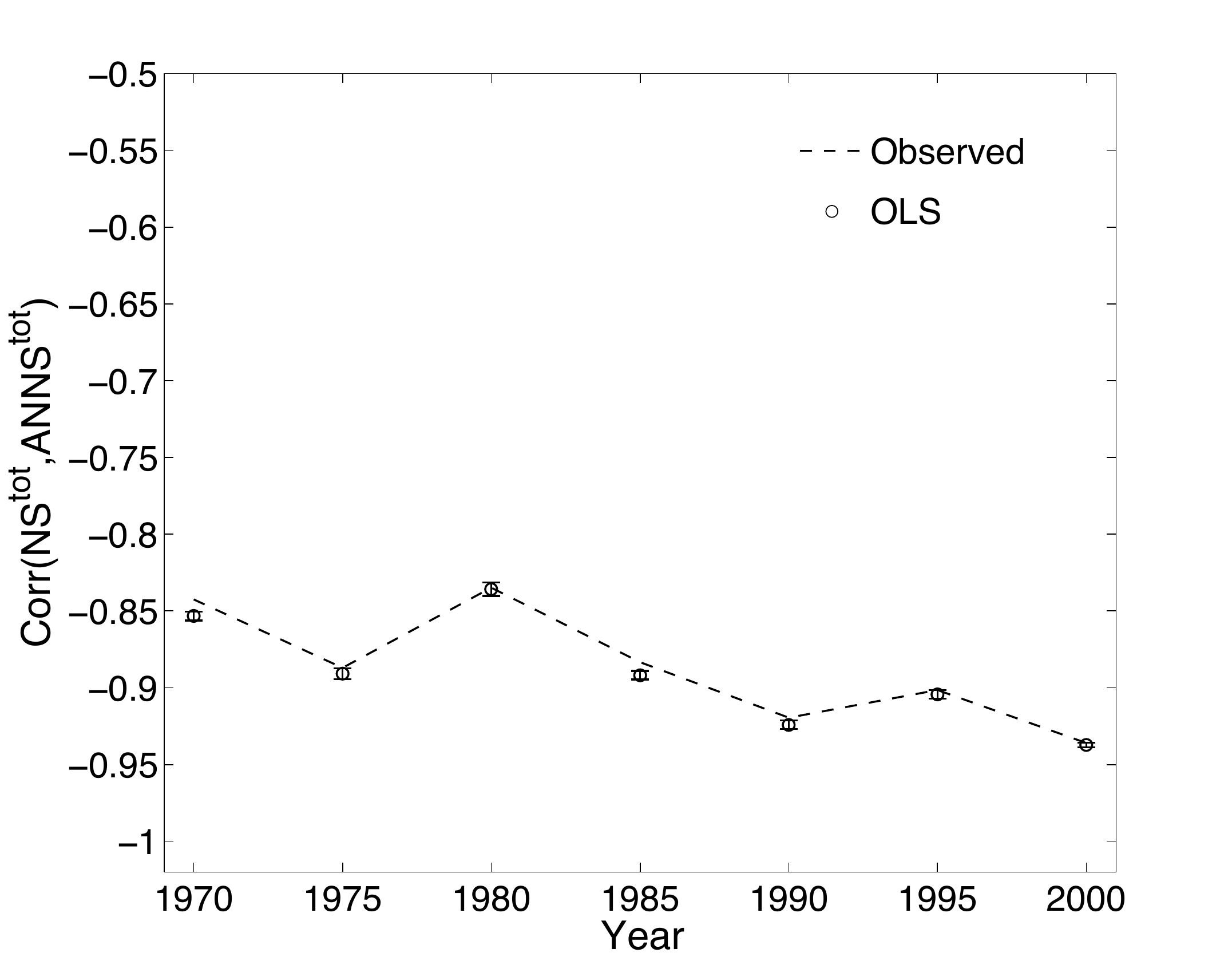}
	\end{minipage}
	\begin{minipage}[t]{5.7cm}
		\includegraphics[width=5.7cm]{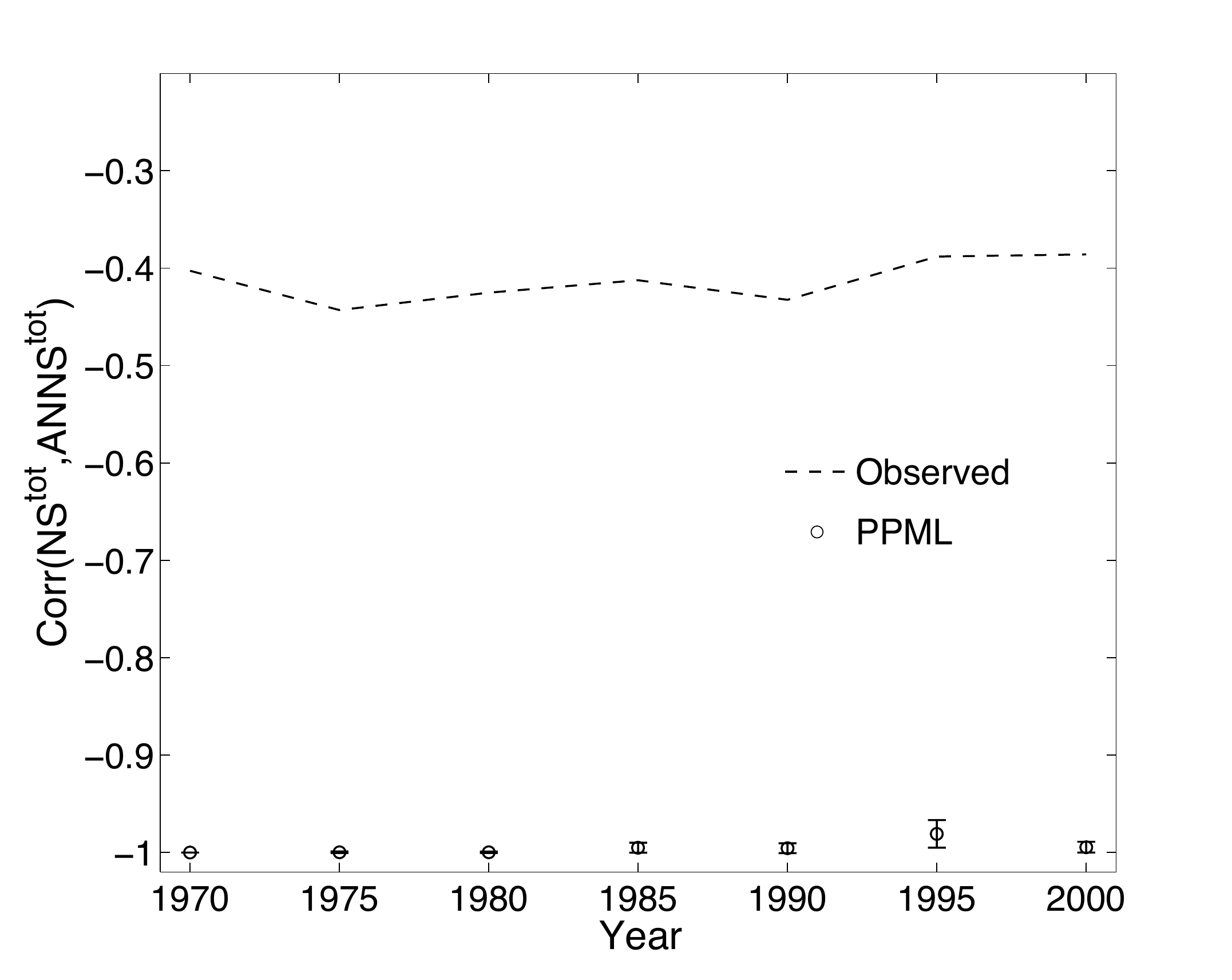}
	\end{minipage}
	\begin{minipage}[t]{5.7cm}
		\includegraphics[width=5.7cm]{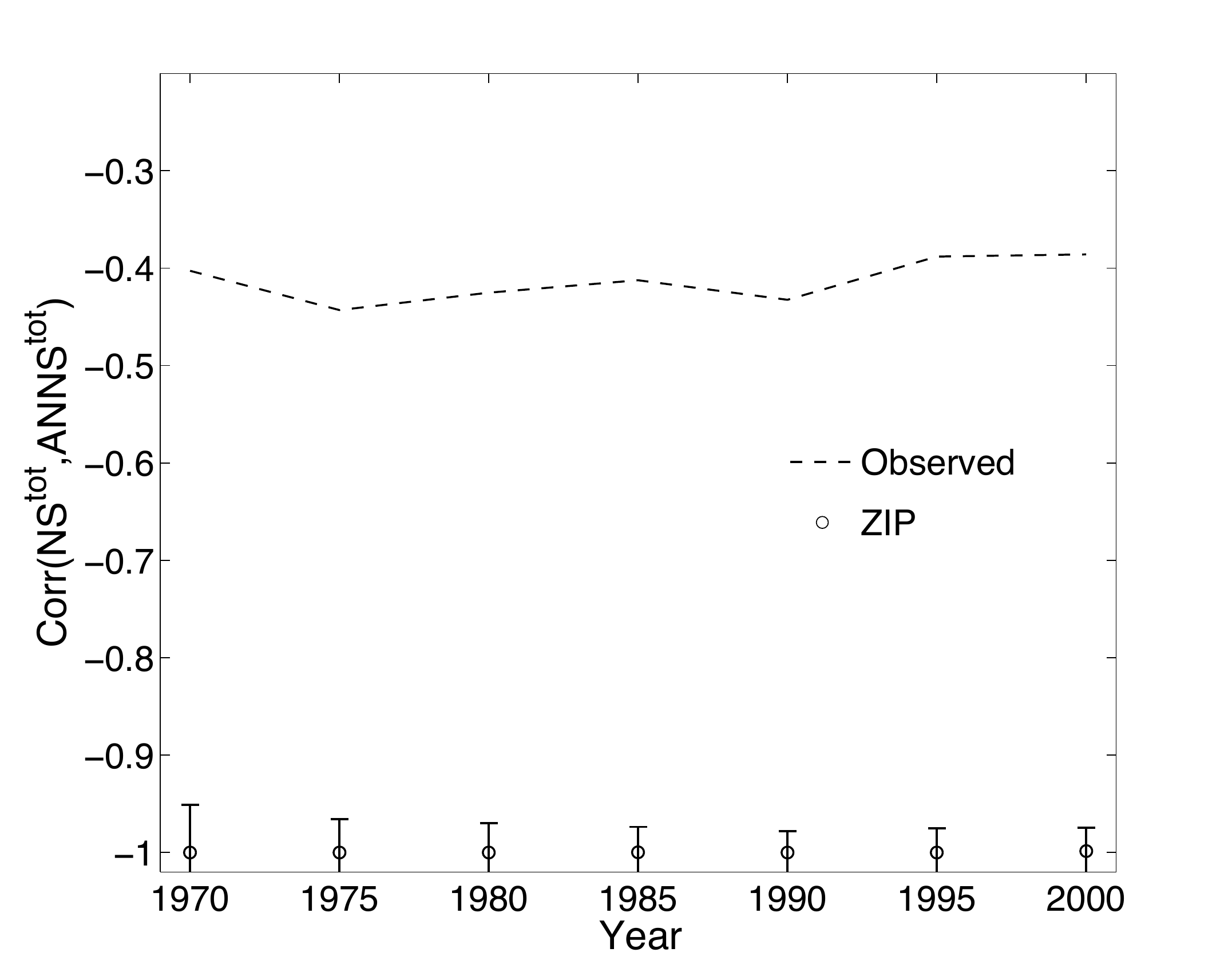}
	\end{minipage}
	\end{center}
	\caption{Observed vs. GM-predicted correlation between total node strenght and total ANNS. 95\% confidence bands are displayed as error bars around predicted values. \textit{Note}: in the OLS plot node strength and ANNS are computed using log of trade flows.} \label{Fig:corr_anns_ts}
\end{figure}

\begin{figure}[t]
	\begin{center}
	\begin{minipage}[t]{5.7cm}
		\includegraphics[width=6cm,height=4.9cm]{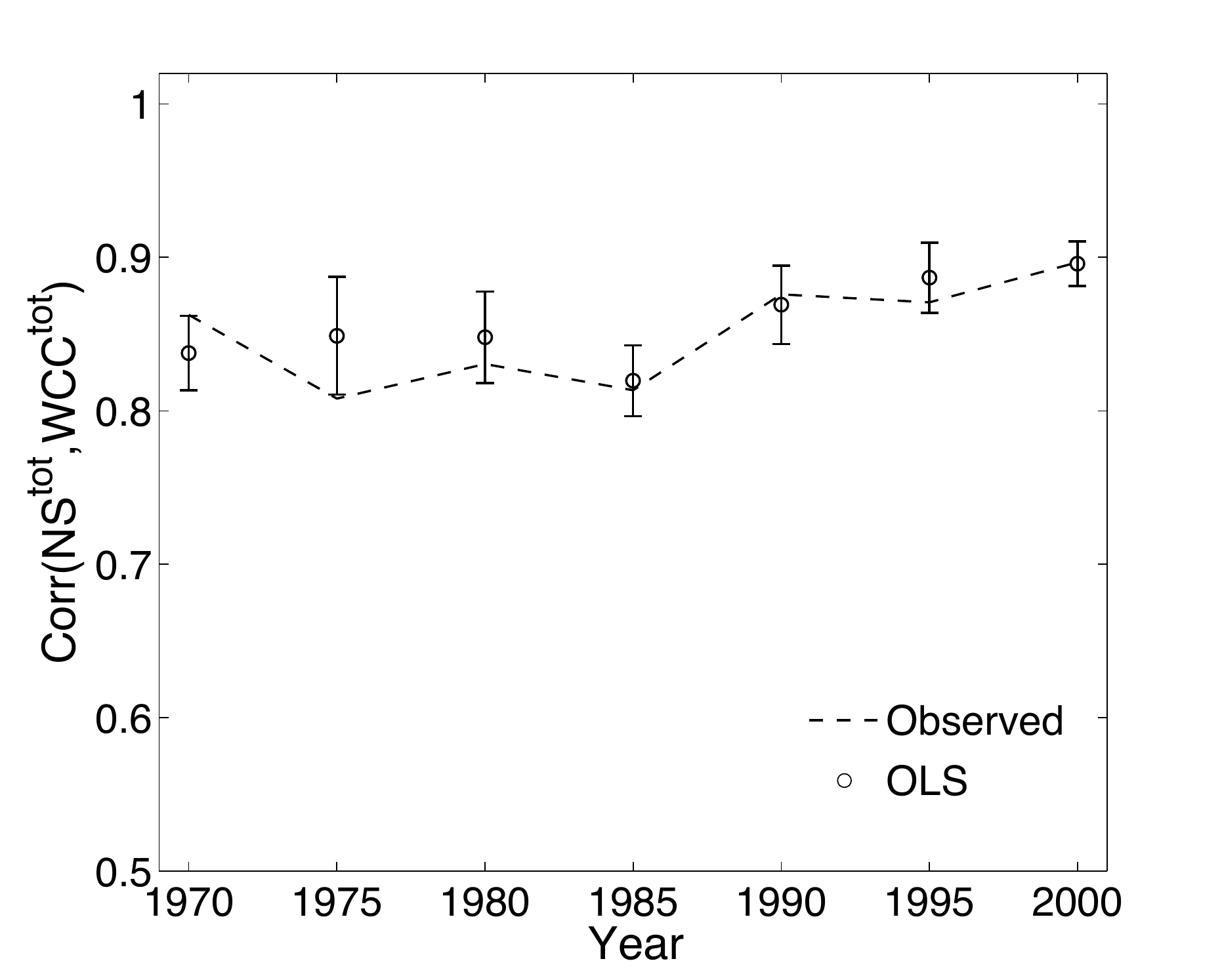}
	\end{minipage}
	\begin{minipage}[t]{5.7cm}
		\includegraphics[width=5.7cm]{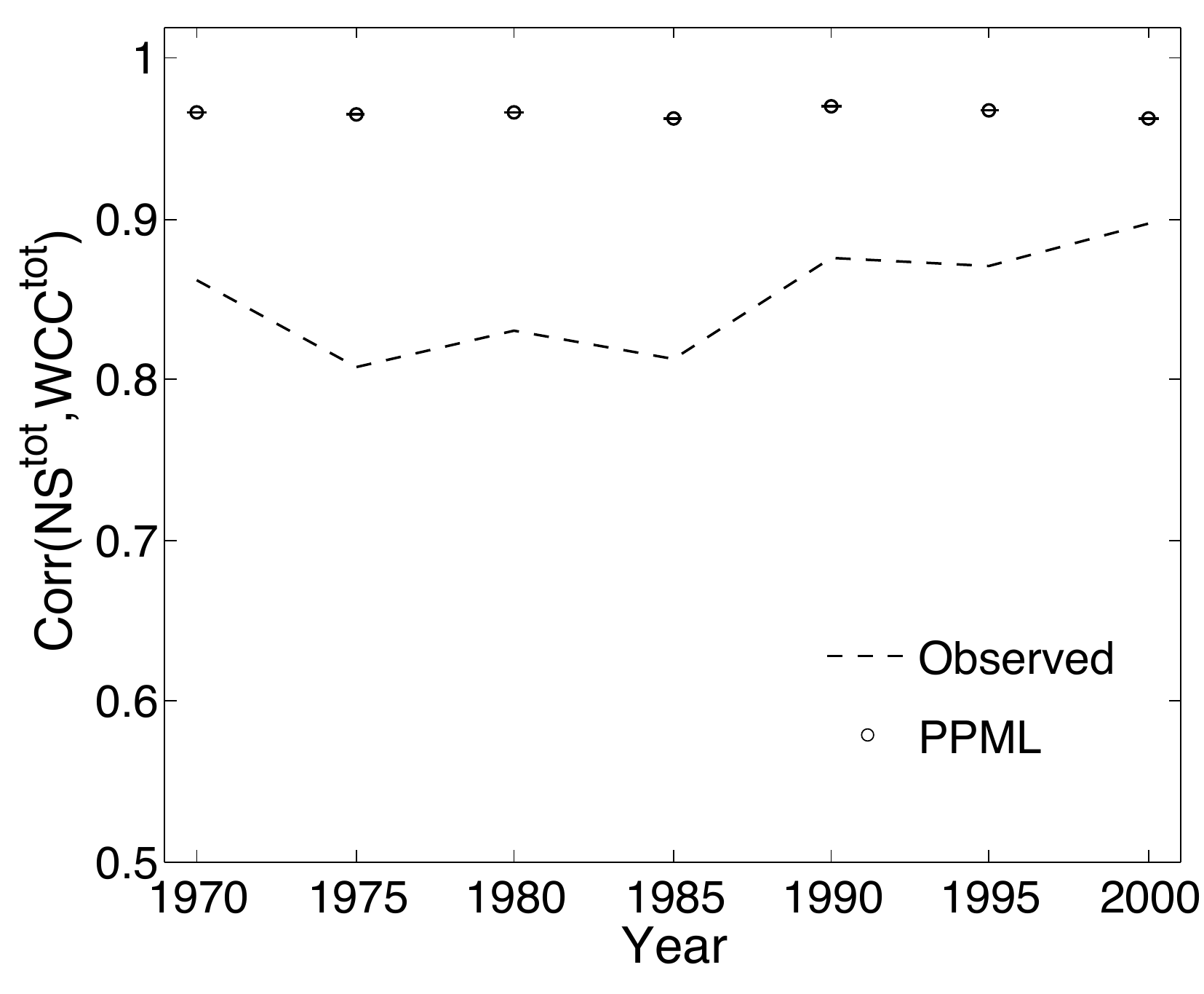}
	\end{minipage}
	\begin{minipage}[t]{5.7cm}
		\includegraphics[width=5.7cm]{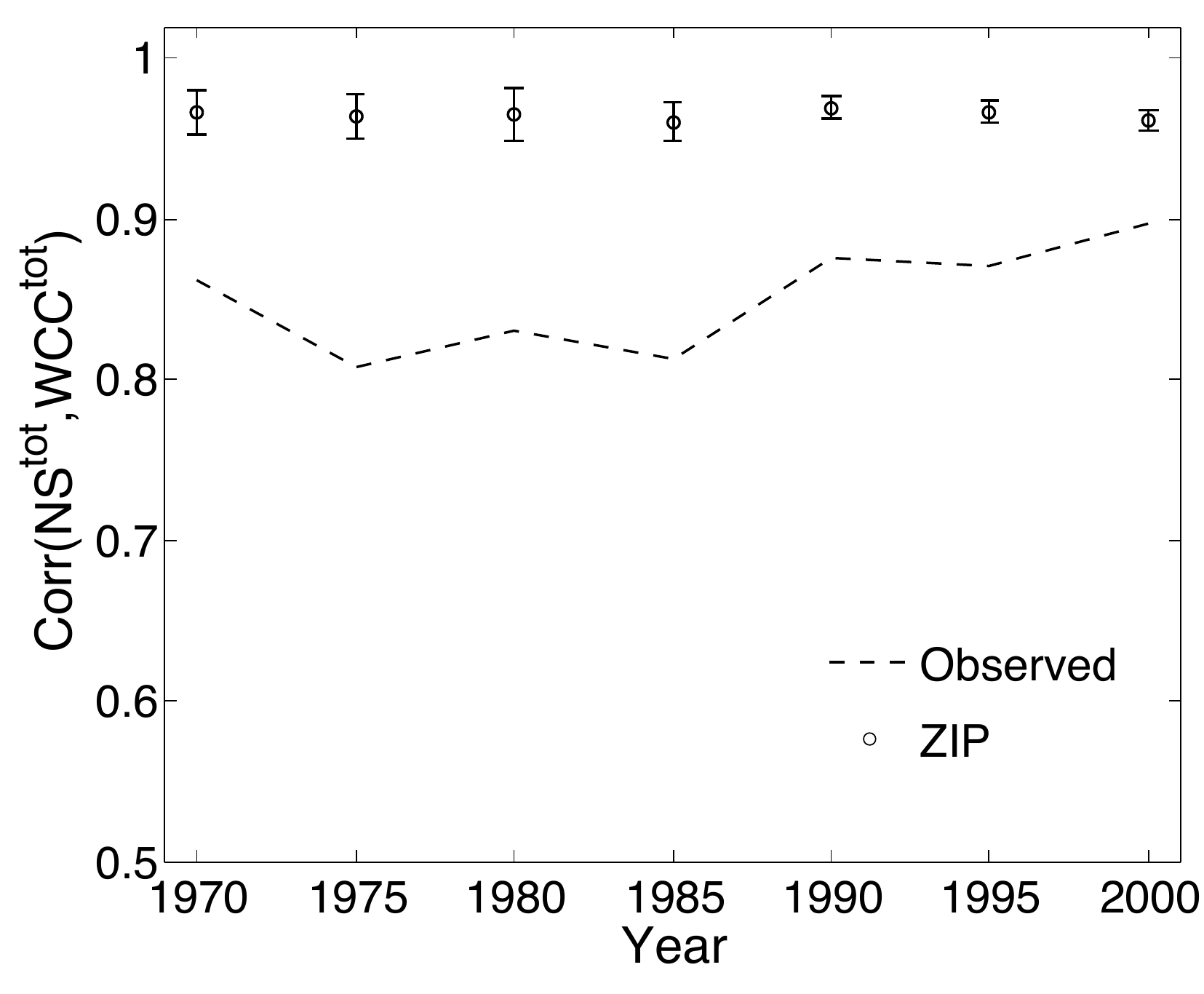}
	\end{minipage}
	\end{center}
	\caption{Observed vs. GM-predicted correlation between total node strenght and total weighted clustering coefficient (WCC). 95\% confidence bands are displayed as error bars around predicted values.}\label{Fig:corr_anns_twcc}
\end{figure}

Another fundamental set of stylized facts characterizing the evolution of the ITN concerns the way in which different network statistics correlate. Figures \ref{Fig:corr_anns_ts} and \ref{Fig:corr_anns_twcc} show observed vs. predicted correlation patterns between, respectively, total ANNS and NS, and total WCC and NS. Note that OLS are able to correctly predict the existing disassortativity emerging between total country trade and average trade of the partners of a node. Conversely, both PPML and ZIP strongly overestimate the magnitude of such negative correlation.\footnote{The difference between observed correlations in the OLS vs PPML/ZIP cases must be attributed to the fact that in OLS we compute all network statistics on logged weight matrices.} OLS estimates are also able to match strength-clustering correlation, even if with less precision than before, as confidence intervals suggest. Again, neither PPML nor ZIP can reproduce the positive strength-clustering correlation characterizing the original weight matrices (in levels), as the underestimation bias appears to be persistently high. 

Correlation results are in line with recent findings by \citet{Squartini_etal_2011a_pre,Squartini_etal_2011b_pre}, who show that higher-order weighted properties in the ITN cannot be reproduced by any random model that takes as given the observed strength sequence (but does not control for the underlying binary structure). Here we show that a satisfactory replication of ITN properties can be achieved only if one fixes the binary structure and attributes link weights using a GM. As soon as the binary structure is badly reproduced, one also looses the possibility to correctly recover weighted-network patterns, primarily because most of weighted-network statistics are inherently dependent on the binary representation.              

So far, we have been studying the performance of GM predictions for weighted undirected statistics. In fact, total strength, ANNS and clustering all neglect the directed nature of trade flows and ensuing asymmetries, as they do not discriminate between in and out links (i.e., import and export flows). To check if the foregoing results also apply in the case of weighted-network \textit{directed} statistics, which instead take fully into account trade-flow directionality, we have studied predicted vs. observed values of population averages of such statistics and their correlation. We have focused on in- and out-strength, and the breakdown in four directed statistics of ANNS and WCC (see Table \ref{table:net_stats}). In all cases, all results obtained above hold. In particular, OLS can easily reproduce all versions of average disaggregated ANNS, while it slightly underestimates the average of all directed clustering coefficients. Both PPML and ZIP fail to capture average ANNS and WCC. Incidentally, according to K-S tests, all three methods are able to fully explain both in- and out-strength distributions, but only OLS predictions get higher-order statistic distributions right. All correlations\footnote{Among all possible correlations of directed statistics with node in- and out-strength we have selected only those economically more relevant. For example, we have focused on the correlation coefficient between $ANNS^{out,in}$ and $NS^{out}$ (and not that between $ANNS^{out,in}$ and $NS^{in}$) because one is much more interested in understanding whether a country that \textit{exports} more, in turn exports to countries that imports more, rather than knowing whether a country that \textit{imports} more, in turn exports to countries that imports more.} are correctly predicted by OLS (with a slight overestimation as far as clustering-strength relationships are concerned), whereas  both PPML and ZIP always fail in replicating correlations among directed statistics. Once again, the ability to predict the binary (directed) structure of the ITN becomes crucial: despite the fact that the three methods correctly replicate the correlation between in- and out-strength, only the OLS (by construction) exploits a perfect prediction of the binary structure, and therefore results in a good approximation of the patterns characterizing weighted statistics.

\subsection{Binary Statistics} \label{SubSec:Binary}
Our weighted-network exercises show that the GM can provide a quite satisfactorily picture of ITN properties only if one restricts the estimation to strictly-positive trade flows, i.e. if the observed binary structure is taken as given. The fact that binary trade links play a crucial role in explaining ITN weighted topology indicates that any GM model aiming at endogenously estimating binary links must somewhat take into account the discrete nature of the binary ITN and try to obtain a more accurate estimation of the exact location of the zeros in trade matrices. 

But is the GM able to correctly predict the binary structure of the ITN? In other words, can one employ the independent variables traditionally used in GM equations to predict whether a trade link exists or not? To address this issue, we employ the most natural candidate model for estimating the probability that a given link is present, i.e. a logit specification. More specifically, for any cross-section $t$, we estimate:
\begin{equation}
Prob\{a_{ij}=1|x_{ij}\}=\frac{\exp \{x_{ij}\cdot \theta\}}{1+\exp \{x_{ij}\cdot \theta\}}=\Lambda(x_{ij};\theta) \label{eq:logit} 
\end{equation}
Notice that Eq. \eqref{eq:logit} is exactly the functional form that we fit in the first stage of the ZIP estimation. Therefore, we can employ first-stage estimates for a zero flow $\hat \psi_{ij}$ from the ZIP model and build the predicted probability matrices $\hat \Xi$, whose generic entry $\hat \xi_{ij}=1-\hat \psi_{ij}$ represents the estimated probability of observing a directed link from country $i$ to country $j$ in that year.\footnote{We have also experimented with the matrix of predicted probabilities coming from the full ZIP estimation, where each element equals $\hat{\psi}_{ij}+(1-\hat{\psi}_{ij})\hat \mu_{ij}$, without noticing any dramatic changes in the results we present below.} 

Of course, as already mentioned, $\hat \xi_{ij}>0$ for all links $ij$. Therefore, if we just employ $\hat \Xi$, we will end up with a full-network for the predicted binary ITN, as it happens for PPML and ZIP procedures. This impairs all subsequent analyses, as they strongly depend on a correct estimation of the binary structure.  

In what follows, we propose a strategy to employ the GM to predict the binary ITN. To get a reasonable prediction for the binary ITN, we proceed in two ways. First, for each year, we take the matrix $\hat \Xi$ and we delete all the links associated to predicted probabilities smaller than the observed ITN density. This generates a predicted binary ITN with a density approximately similar (for numerical reasons) to the observed one.\footnote{We also performed two alternative threshold-based exercises. In the first one, we delete all the links associated to predicted probabilities smaller than a given threshold $s$, which is chosen so as to approximately match the empirically-observed ITN density. In the second one, the optimal threshold is chosen so as to minimize the Manhattan distance between the observed adjacency matrix and the predicted binary one, where the latter is defined, for each given threshold $s$, as the binary matrix where a link is in place if and only if $\hat \xi_{ij}<s$. Both procedures lead to very similar optimal thresholds, which are in turn very close to the density we get by straightforwardly setting the threshold equal to $\rho$. Therefore, we present here only results for the case where the threshold is equal to observed density.} This leads to the following:

\begin{Def}[Density-Induced Predicted Binary ITN]
The density-induced predicted binary International Trade Network, for each given cross-section $t$, is represented by a binary-directed graph, where the nodes are countries and the adjacency asymmetric matrix $\hat A (\rho)$ has entries $\hat a_{ij}(\rho)=1$ if and only if $\hat \xi_{ij}>\rho$, where $\rho$ is observed ITN density in year $t$.
\end{Def}

Second, we exploit a simulation-based procedure that, instead of using density-matching thresholds, fully exploits the information in $\hat \Xi$. More precisely, in each year, we generate a sample of $M$ independent adjacency matrices $\hat A^m=\{\hat a_{ij}^m\}$, for $m=1,\dots,M$ where in each sample $\hat a_{ij}^m$ is drawn from a Bernoulli distribution with parameter $\hat \xi_{ij}$, independently across all pairs $ij$. More formally:

\begin{Def}[Bernoulli Predicted Binary ITN]
The Bernoulli predicted binary International Trade Network, for each given cross-section $t$, is represented by a distribution of $M$ binary-directed graphs, where for each graph the nodes are countries and the adjacency asymmetric matrix $\hat A^m$, $m=1,\dots,M$, has entries $\hat a_{ij}^m$ that are drawn independently from a Bernoulli distribution with parameter $\hat \xi_{ij}$.
\end{Def}
In our exercises, we set $M=10,000$ and we employ simulated predicted matrices to compute Monte-Carlo standard deviations for all statistics of interest (of course with the first method we do not have any source of variation in predicted values, therefore no error bars can be computed). Note also that density-induced predicted binary ITN preserves almost exactly observed density, while the Bernoulli predicted binary ITN preserves that quantity only on average.  

\begin{figure}[t]
	\begin{center}
	\begin{minipage}[t]{5.7cm}
		\includegraphics[width=6cm,height=4.9cm]{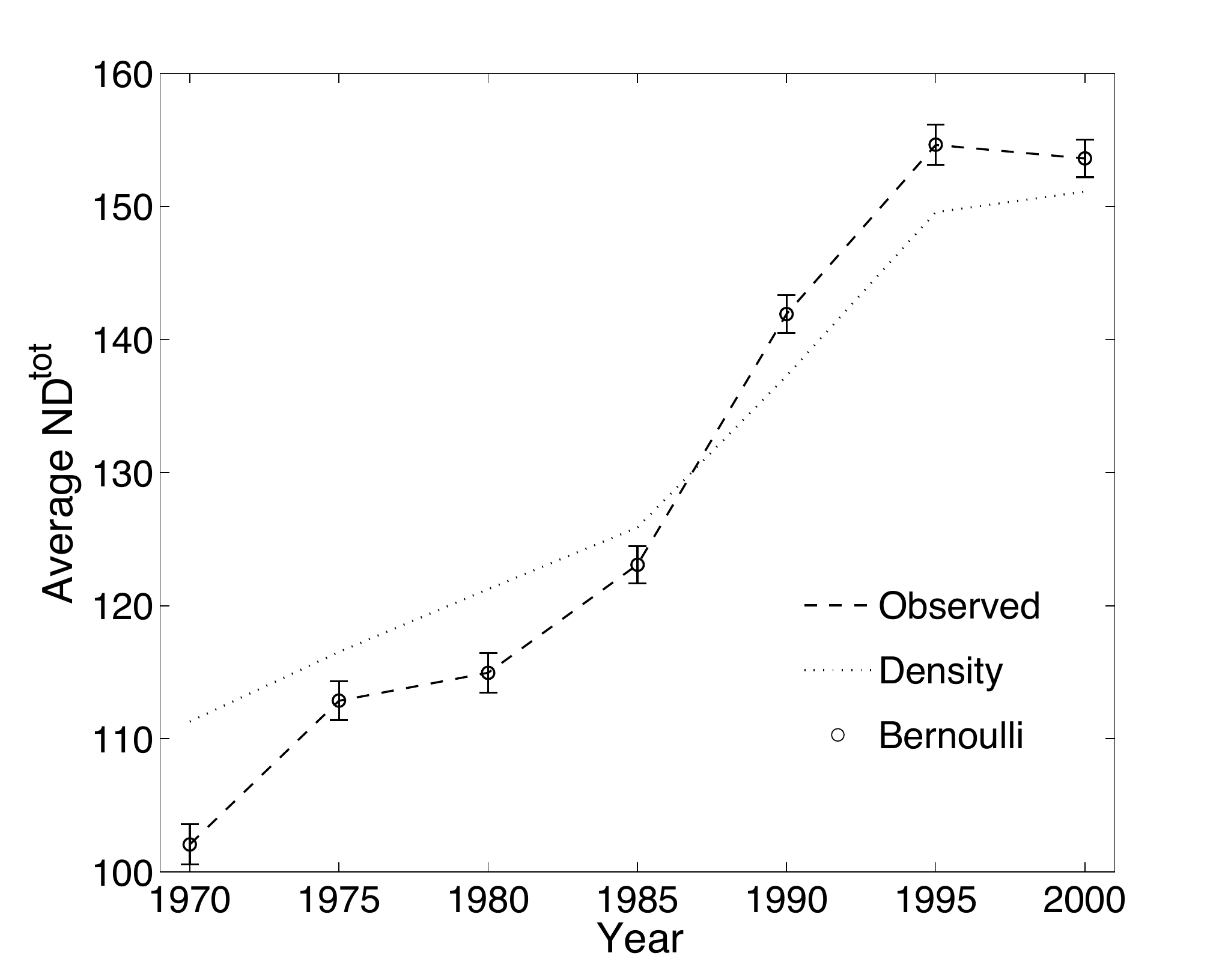}
	\end{minipage}
	\begin{minipage}[t]{5.7cm}
		\includegraphics[width=5.7cm,height=4.9cm]{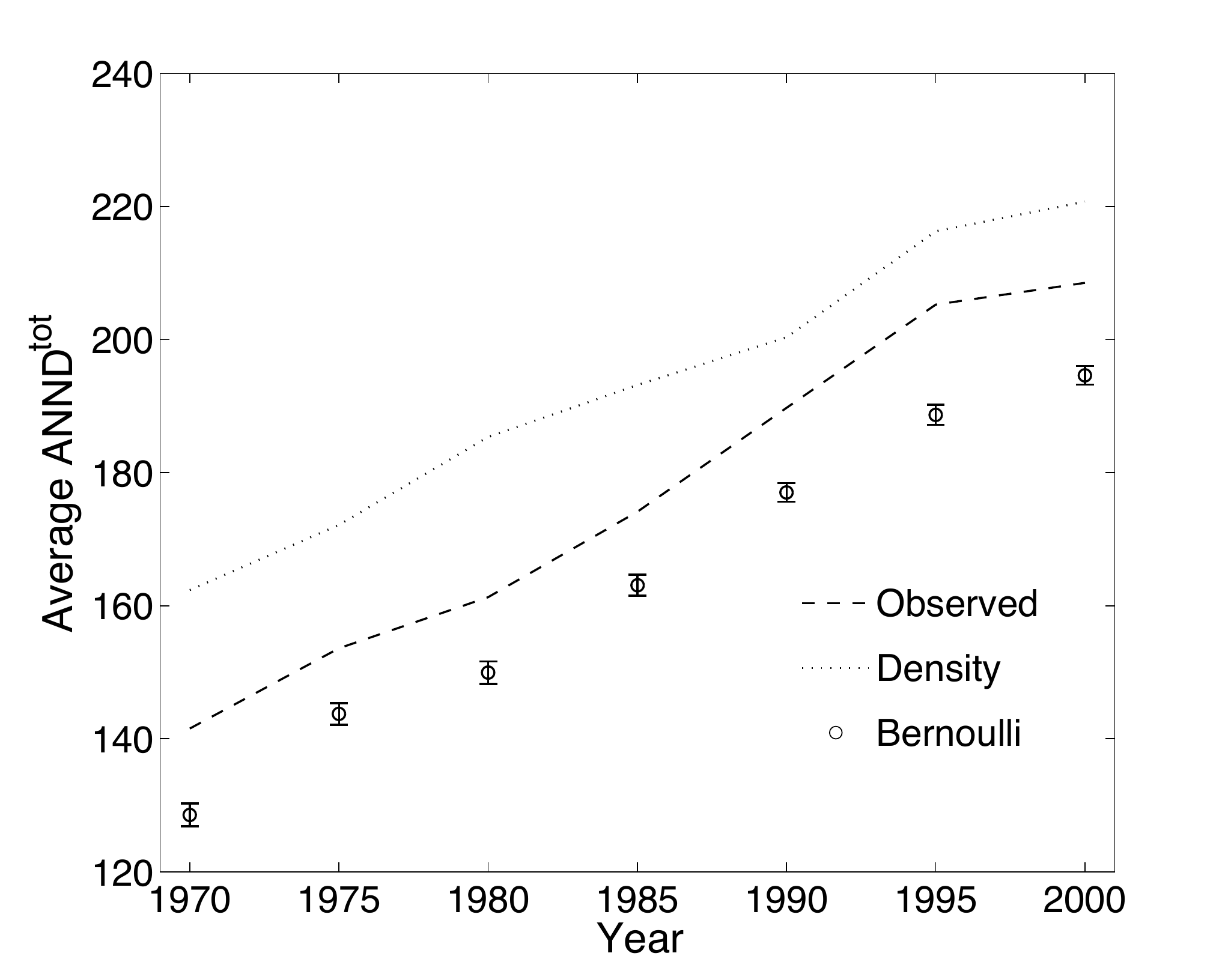}
	\end{minipage}
	\begin{minipage}[t]{5.7cm}
		\includegraphics[width=5.7cm,height=4.9cm]{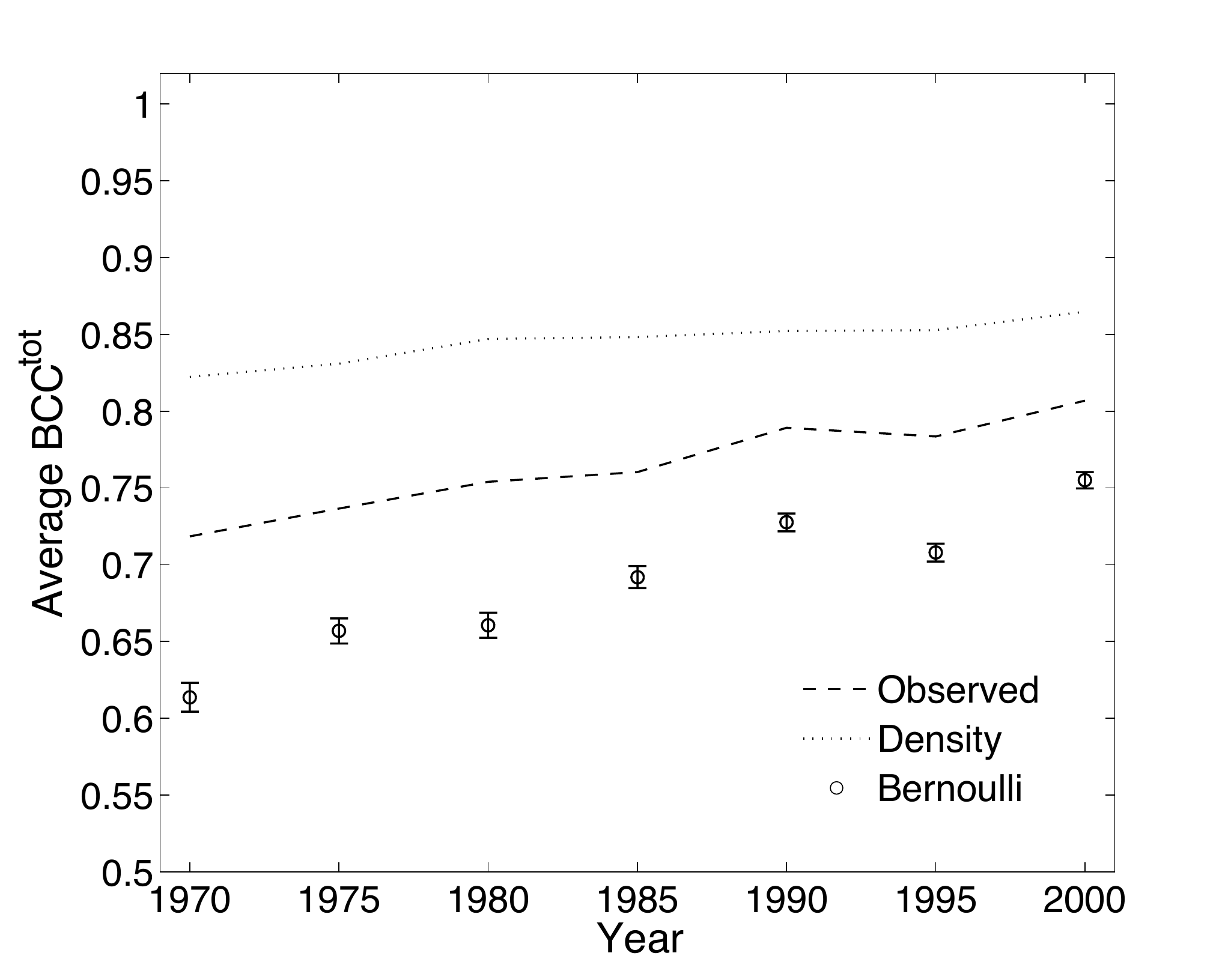}
	\end{minipage}
	\end{center}
	\caption{Observed vs. GM-predicted average statistics in the binary ITN. Logit estimation. Density: average statistics in the Density-Induced Predicted Binary ITN (see Definition 4). Bernoulli: average statistics in the Bernoulli Predicted Binary ITN (see Definition 5). 95\% confidence bands are displayed as error bars around predicted Bernoulli values.}\label{Fig:ave_binary}
\end{figure}

\begin{figure}[t]
	\begin{center}
	\begin{minipage}[t]{8cm}
		\includegraphics[width=8cm]{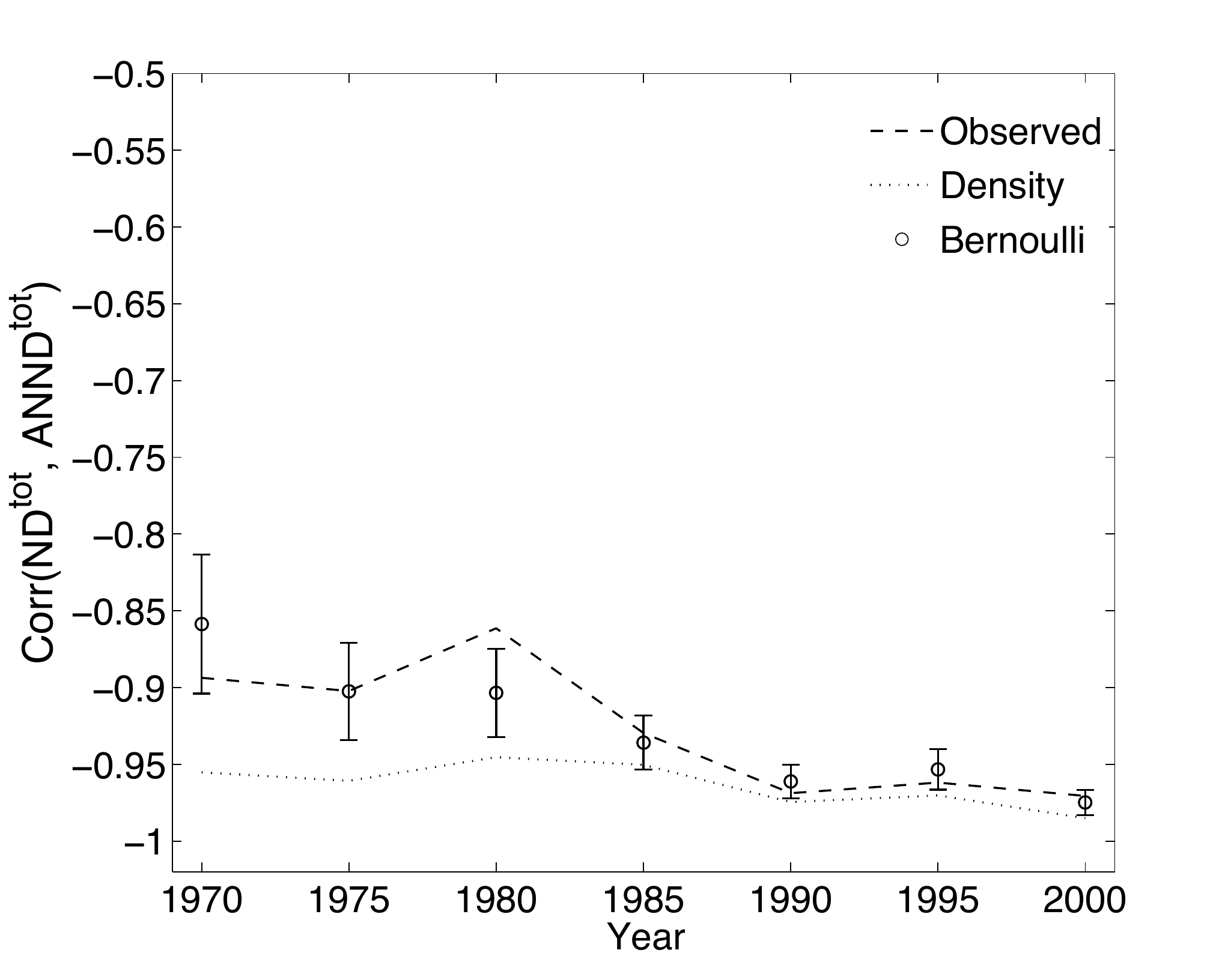}
	\end{minipage}
	\begin{minipage}[t]{8cm}
		\includegraphics[width=8cm]{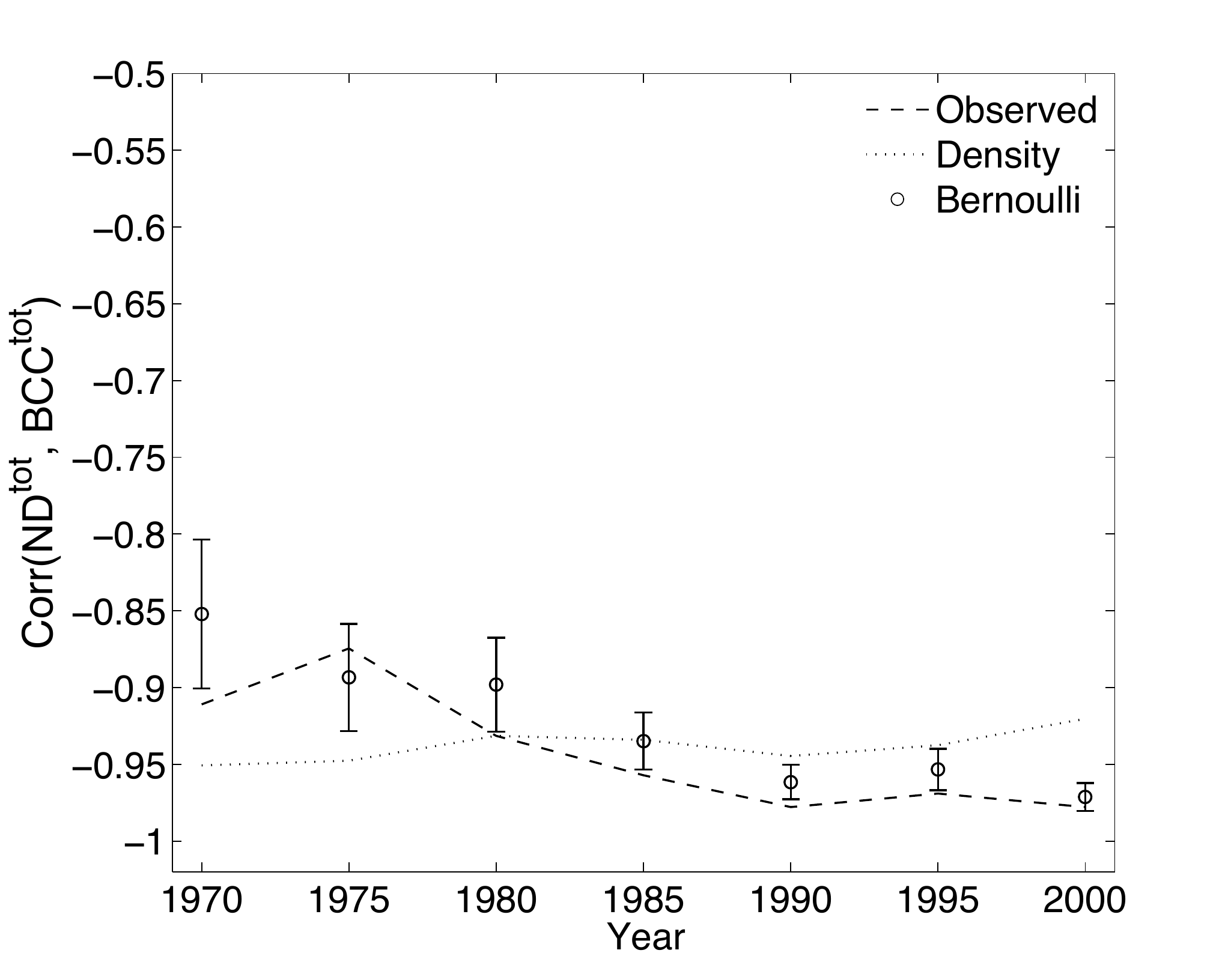}
	\end{minipage}
	\end{center}
	\caption{Observed vs. GM-predicted correlation between statistics in the binary ITN. Logit estimation. Density: average statistics in the Density-Induced Predicted Binary ITN (see Definition 4). Bernoulli: average statistics in the Bernoulli Predicted Binary ITN (see Definition 5). 95\% confidence bands are displayed as error bars around predicted Bernoulli values.}\label{Fig:corr_binary}
\end{figure}

Our main results are reported in Figures \ref{Fig:ave_binary} and \ref{Fig:corr_binary}, where we plot observed binary statistics vs. predicted ones, using density-induced and Bernoulli procedures (see Definitions 4 and 5 above).\footnote{We focus here only on undirected measures. All main results hold also for directed network statistics.} To begin with, note that both density-induced and Bernoulli predictions are quite successful in tracking average total ND. Actually, Bernoulli-predicted binary ITN can exactly replicate, on average, that statistics. Conversely, density-induced predictions slightly deviate from observed values.\footnote{This is entirely due to numerical problems, as often it is impossible to purge from the predicted matrix $\hat \Xi$ a number of links so as to perfectly match observed density.} This is not surprising, as it means that both models are able to get a good proxy of observed density.\footnote{Average total ND is indeed equal to $L/N$, where $L$ is the number of links, and hence equals $(N-1)\rho$.} 

The fact that a Logit estimation is on average able to predict observed density explains why a ZIP model, which employs the very same Logit specification in its first stage, predicts very well average total NS. For that statistics is an average over all existing links and it is not so much affected by where these links are actually located. This is not true of ANNS and WCC, which in fact are badly reproduced by a ZIP model because require a more precise knowledge of where links are placed.

A similar problem arises in the binary ITN for both density-induced and Bernoulli predictions: the former persistently overestimates observed average ANND and BCC, whereas the latter persistently underestimates them. Again, this hints to an inherent inability of the GM to well predict the presence a link. 

Things seem to improve a bit when we move to correlation structure. Density-induced predictions are able to well capture binary disassortativity in the last part of the sample, but they only partially get right clustering-degree correlation. Conversely, Bernoulli-based predictions seem to perform quite satisfactorily in both cases: although on average observed correlations are rarely replicated, the inherent variability of this procedure allows one to conclude that there exists a sufficiently large number of simulations where predicted correlations are very similar to observed ones. 

\section{Concluding Remarks\label{Section:Conclusions}}
In this paper, we have studied whether a gravity model (GM), the work-horse theoretical reference in international trade, can explain the statistical properties of the international-trade network. 

Our exercises show that the GM does a very good job in replicating the weighted-network structure of the ITN only if one fixes its binary architecture equal to the observed one. More generally, the GM performs very badly when asked to predict the presence of a link, or the level of the trade flow it carries, whenever the binary structure must be simultaneously estimated.

Therefore, the GM turns out to be a good model for estimating trade flows, but not to explain why a link in the ITN gets formed and persists over time. In other words, knowing country-specific variables (country GDP, etc.) and country bilateral interactions (bordering conditions, belonging to the same RTA, etc.) is not enough to predict the presence of a link. However, conditional on the information that a link exists, such variables can well predict how much trade that link actually carries.

Notice that these results are largely independent on which variables are actually entering the gravity equation we fit to the data. In the foregoing exercises, we have used a standard specification where many of the most-employed GM variables enter the regression. We have also tried and augment the equation with other explanatory variables that resulted statistically not significant, but can nevertheless improve the percentage of explained trade-flow variance, without observing any dramatic increase in the goodness of fit of ITN network statistics. 

In order to better explain the topological properties of the ITN many alternative strategies may be pursued. First, one may consider to augment a GM specification with network-related variables. It may be indeed argued that if standard economic variables entering in the GM are not enough to explain link formation, perhaps this is because the presence of a link between any two countries might be actually explained by the very local structure of the network (e.g., degrees of the two countries, etc.). Of course this introduces some endogeneity to the problem, because the presence of a link in turn affects local network properties. By properly dealing with endogeneity issues in estimation, one can hope to better explain the binary structure of the ITN.

Second, one might borrow social-network statistical methodologies currently employed to model the evolution of directed graphs over time as continuous-time Markov processes \citep{Snijders_2005}. For example, one may envisage setups where each single node chooses its outgoing link (i.e. whether to export to another country or not) based on a myopic optimization of some objective function, where the latter may be the result of many firm-level decisions within the origin country.

Finally, one may think to explore international-trade models where the decision of a firm located in country A to export goods to country B, which possibly never imported products from A before, is rooted in a more detailed micro-foundation. This may require to blend together two strands of literature, one on the role of heterogeneous firms in international trade \citep{Melitz_2003,Bernard_etal_2007} and the other on models of trade network formation based on simple aggregate dynamics \citep{Garla2004,Bhatta2007a,Riccaboni_Schiavo_2010}.

\singlespacing
\bigskip



\newpage


\begin{sidewaystable}[htbp]
\begin{center}
\begin{tabular}{p{8cm}rrrrrrr}  \hline
 & 1970 & 1975 & 1980 & 1985 & 1990 & 1995 & 2000 \\ \hline
Countries (No.) & 129 & 135 & 142 & 148 & 145 & 157 & 154 \\
Trade Flows (No.) & 6583 & 7618 & 8162 & 9108 & 10289 & 12138 & 11828 \\
Density & 0.40 & 0.42 & 0.41 & 0.42 & 0.49 & 0.50 & 0.50 \\
Average Trade & 51.03 & 56.43 & 57.48 & 61.54 & 70.96 & 77.31 & 79.81 \\
Countries making up 50\% of trade & 7 & 8 & 7 & 7 & 7 & 8 & 8 \\
Flows making up 50\% of trade & 73 & 99 & 90 & 73 & 69 & 74 & 79 \\
Countries making up 90\% of trade & 39 & 39 & 38 & 37 & 31 & 32 & 33 \\
Flows making up 90\% of trade & 794 & 900 & 894 & 871 & 749 & 826 & 855 \\
\% Countries making up 50\% of trade & 5.43\% & 5.93\% & 4.93\% & 4.73\% & 4.83\% & 5.10\% & 5.19\% \\
\% Flows making up 50\% of trade & 1.11\% & 1.30\% & 1.10\% & 0.80\% & 0.67\% & 0.61\% & 0.67\% \\
\% Countries making up 90\% of trade & 30.23\% & 28.89\% & 26.76\% & 25.00\% & 21.38\% & 20.38\% & 21.43\% \\
\% Flows making up 90\% of trade & 12.06\% & 11.81\% & 10.95\% & 9.56\% & 7.28\% & 6.81\% & 7.23\% \\
\hline 
\end{tabular}
\caption{\citet{Subramanian_Wei_2003_database} Database. Summary statistics.} \label{Tab:Summary_Stats}
\end{center}
\end{sidewaystable}

\begin{table}
\begin{small}
\centering
\begin{tabular}{l l p{5cm} l}
\hline
Label        & Related to & Description & Source \\
\hline
$W$          & Link       & Imports in U.S. Dollars & \cite{Subramanian_Wei_2003_database} \\
$Y$          & Country    & Gross-domestic product & \cite{Subramanian_Wei_2003_database} \\
area         & Country    & Country area in $\text{Km}^2$ & \cite{Subramanian_Wei_2003_database} \\
pop          & Country    & Country population & \cite{Subramanian_Wei_2003_database} \\
$d$          & Link       & distance between two countries, based on bilateral distances between the largest cities of those two countries, weighted by the share of the city in the overall countryÕs population & CEPII (http://www.cepii.fr/) \\
landl        & Country    & Dummy variable equal to 1 for landlocked Countries & CEPII (http://www.cepii.fr/) \\
continent    & Country    & Categorical variable indicating the continent of the country  & CEPII (http://www.cepii.fr/) \\
contig       & Link       & Contiguity dummy equal to 1 if two countries share a common border & CEPII (http://www.cepii.fr/) \\
comlang\_off & Link       & Dummy equal to 1 if both countries share a common official language & CEPII (http://www.cepii.fr/) \\
comcol       & Link       & Dummy equal to 1 if both countries have had a common colonizer & CEPII (http://www.cepii.fr/) \\
colony       & Link       & Dummy equal to 1 if both countries have ever had a colonial link & CEPII (http://www.cepii.fr/) \\
curcol       & Link       & Dummy equal to 1 if both countries are currently in a colonial relationship & CEPII (http://www.cepii.fr/) \\
comrelig     & Link       & Percentage in which both countries share religions & CEPII (http://www.cepii.fr/) \\
comcur       & Link       & Dummy equal to 1 if both countries have a currency unions & CEPII (http://www.cepii.fr/) \\
gsp          & Link       & Dummy equal to 1 if both countries share a generalized system of preferences & CEPII (http://www.cepii.fr/) \\
rta          & Link       & Dummy variable equal to 1 if both countries involved in regional, bilateral or preferential trade agreements & WTO (http://www.wto.org/) \\
\hline
\end{tabular}
\caption{Variables employed in the gravity-model estimation.}\label{table:vars}
\end{small}
\end{table}

\begin{table}
\begin{small}
\centering
\begin{tabular}{l r r r r r }
\hline
\hline
Regressor &  & OLS(W$>$0) & PPML & ZIP-Poisson & ZIP-Logit\\
\hline
ln\_gdp\_i & $(\alpha_1)$ & 1.415***(0.023) & 1.302***(0.049) & 1.278***(0.048) & -0.954***(0.025)\\
ln\_gdp\_j & $(\alpha_2)$ & 1.323***(0.021) & 1.697***(0.047) & 1.65***(0.047) & -0.961***(0.023)\\
ln\_dist\_ij & $(\alpha_3)$ & -1.034***(0.023) & -0.725***(0.033) & -0.721***(0.033) & 0.533***(0.033)\\
ln\_area\_i & $(\beta_{11})$ & -0.068***(0.013) & -0.097***(0.022) & -0.09***(0.022) & 0.107***(0.013)\\
ln\_area\_j & $(\beta_{21})$ & -0.108***(0.012) & -0.14***(0.035) & -0.135***(0.034) & 0.15***(0.013)\\
ln\_pop\_i & $(\beta_{12})$ & -0.402***(0.025) & -0.401***(0.07) & -0.393***(0.068) & 0.206***(0.026)\\
ln\_pop\_j & $(\beta_{22})$ & -0.42***(0.025) & -0.773***(0.057) & -0.744***(0.056) & 0.25***(0.026)\\
landl\_ci & $(\delta_{11})$ & -0.456***(0.051) & -0.509***(0.092) & -0.48***(0.091) & 0.517***(0.047)\\
landl\_cj & $(\delta_{21})$ & -0.472***(0.046) & -0.451***(0.124) & -0.426***(0.123) & 0.579***(0.047)\\
continent\_i & $(\delta_{12})$ & 0.004(0.02) & -0.16***(0.038) & -0.153***(0.038) & 0.043*(0.019)\\
continent\_j & $(\delta_{22})$ & -0.05**(0.019) & -0.257***(0.042) & -0.259***(0.041) & -0.086***(0.018)\\
contig & $(\theta_1)$ & 0.823***(0.112) & 0.572***(0.113) & 0.622***(0.114) & 0.996***(0.216)\\
comlang\_off & $(\theta_2)$ & 0.637***(0.055) & 0.407***(0.084) & 0.376***(0.083) & -0.728***(0.061)\\
comcol & $(\theta_3)$ & 0.785***(0.084) & 0.399(0.276) & 0.41(0.274) & -0.26**(0.077)\\
colony & $(\theta_4)$ & 1.091***(0.088) & -0.252**(0.093) & -0.226**(0.092) & 0.476*(0.239)\\
curcol & $(\theta_5)$ & -2.334(1.923) & 0.156(0.737) & 0.345(0.718) & 2.081(1.291)\\
comrelig & $(\theta_6)$ & 0.266***(0.066) & -0.094(0.109) & -0.148(0.109) & -0.676***(0.081)\\
comcur & $(\theta_7)$ & 0.554***(0.111) & -0.139(0.107) & -0.136(0.107) & -1.493***(0.18)\\
gsp & $(\theta_8)$ & 0.484***(0.047) & 0.349***(0.107) & 0.303***(0.106) & -2.015***(0.116)\\
rta & $(\theta_9)$ & 0.338***(0.053) & 0.204**(0.078) & 0.181**(0.078) & -0.672***(0.078)\\
\_cons & $(\gamma)$ & -20.666***(0.394) & -21.693***(0.852) & -20.8***(0.868) & 21.636***(0.487)\\
\hline
No. Obs &  & 11828 & 23562 & 11828 & 23562\\
F or Wald chi2 &  & 1423 & 15029 & 14083 & 5691.15\\
Prob $>$ F or chi2  &  & 0 & 0 & 0 & 0\\
R2  or Pseudo R2  &  & 0.68 & 0.93 & 0.92 & 0.43\\
Vuong Z &  & -  & -  & 82.76 & \\
Prob $>$ Z &  & -  & -  & 0 & \\
\hline
\hline
\end{tabular}
\caption{GM estimation. Year: 2000. ZIP-Poisson: second stage of the ZIP estimation process. ZIP-Logit: first stage of the ZIP estimation process. \textit{Note}: $i$ = exporter; $j$ = importer.}\label{table:gm_est}
\end{small}
\end{table}

\newpage

\begin{sidewaystable}[h]
\centering
\begin{tabular}{p{4cm}|p{8cm}|p{8cm}}
\textbf{Topological \newline Properties} &\ \newline \textbf{Binary} & \ \newline \textbf{Weighted}\\
\hline
Degrees/Strengths & $ND_{i}^{in}=k_{i}^{in}=\sum_{j}a_{ji}\newline ND_{i}^{out}=k_{i}^{out}=\sum_{j}a_{ij}\newline ND_{i}^{out}=k_{i}^{tot}=k_{i}^{in}+k_{i}^{out}$ & $NS_{i}^{in}=s_{i}^{in}=\sum_{j}w_{ji}\newline NS_{i}^{out}=s_{i}^{out}=\sum_{j}w_{ij}\newline NS_{i}^{tot}=s_{i}^{tot}=s_{i}^{in}+s_{i}^{out}$\\
\hline
ANND/ANNS & $ANND_{in,in}=\frac{\sum_{j}a_{ji}k_{j}^{in}}{k_{i}^{in}}$ \newline $ANND_{in,out}=\frac{\sum_{j}a_{ji}k_{j}^{out}}{k_{i}^{in}}$ \newline $ANND_{out,in}=\frac{\sum_{j}a_{ij}k_{j}^{in}}{k_{i}^{out}}$ \newline $ANND_{out,out}=\frac{\sum_{j}a_{ij}k_{j}^{out}}{k_{i}^{out}}$ \newline $ANND^{tot}=\frac{\sum_{j}(a_{ij}+a_{ji})k_{j}^{tot}}{k_{i}^{tot}}$ & $ANNS_{in,in}=\frac{\sum_{j}a_{ji}s_{j}^{in}}{k_{i}^{in}}$ \newline $ANNS_{in,out}=\frac{\sum_{j}a_{ji}s_{j}^{out}}{k_{i}^{in}}$ \newline $ANNS_{out,in}=\frac{\sum_{j}a_{ij}s_{j}^{in}}{k_{i}^{out}}$ \newline $ANNS_{out,out}=\frac{\sum_{j}a_{ij}s_{j}^{out}}{k_{i}^{out}}$ \newline $ANNS^{tot}=\frac{\sum_{j}(a_{ij}+a_{ji})s_{j}^{tot}}{k_{i}^{tot}}$\\
\hline
Clustering & $BCC_{i}^{cyc}=\frac{\sum_{j}\sum_{k}a_{ij}a_{jk}a_{ki}}{k_{i}^{in}k_{i}^{out}-k_{i}^{\leftrightarrow}}$ \newline  $BCC_{i}^{mid}=\frac{\sum_{j}\sum_{k}a_{ik}a_{ji}a_{jk}}{k_{i}^{in}k_{i}^{out}-k_{i}^{\leftrightarrow}}$ \newline  $BCC_{i}^{in}=\frac{\sum_{j}\sum_{k}a_{ki}a_{ji}a_{jk}}{k_{i}^{in}(k_{i}^{in}-1)}$ \newline  $BCC_{i}^{out}=\frac{\sum_{j}\sum_{k}a_{ik}a_{jk}a_{ij}}{k_{i}^{out}(k_{i}^{out}-1)}$ \newline $BCC_{i}^{tot}=\frac{\sum_{j}\sum_{k}(a_{ij}+a_{ji})(a_{jk}+a_{kj})(a_{ki}+a_{ik})}{2\big[k_{i}^{tot}(k_{i}^{tot}-1)-2 k_{i}^{\leftrightarrow}\big]}$ & $WCC_{i}^{cyc}=\frac{\sum_{j}\sum_{k}w_{ij}^{1/3}w_{jk}^{1/3}w_{ki}^{1/3}}{k_{i}^{in}k_{i}^{out}-k_{i}^{\leftrightarrow}}$  \newline  $WCC_{i}^{mid}=\frac{\sum_{j}\sum_{k}w_{ik}^{1/3}w_{jk}^{1/3}w_{ji}^{1/3}}{k_{i}^{in}k_{i}^{out}-k_{i}^{\leftrightarrow}}$ \newline $WCC_{i}^{in}=\frac{\sum_{j}\sum_{k}w_{jk}^{1/3}w_{ji}^{1/3}w_{ki}^{1/3}}{k_{i}^{in}(k_{i}^{in}-1)}$  \newline  $WCC_{i}^{out}=\frac{\sum_{j}\sum_{k}w_{ik}^{1/3}w_{ij}^{1/3}w_{jk}^{1/3}}{k_{i}^{out}(k_{i}^{out}-1)}$  \newline $WCC_{i}^{tot}=\frac{\sum_{j}\sum_{k}(w_{ij}^{1/3}+w_{ji}^{1/3})(w_{jk}^{1/3}+w_{kj}^{1/3})(w_{ki}^{1/3}+w_{ik}^{1/3})}{2\big[k_{i}^{tot}(k_{i}^{tot}-1)-2 k_{i}^{\leftrightarrow}\big]}$\\
\hline
\end{tabular}
\caption{Binary and weighted topological properties of the ITN. \textit{Note}: Time labels are suppressed for notational convenience.}\label{table:net_stats}
\end{sidewaystable}

\begin{table}
\begin{footnotesize}
\centering
\begin{tabular}{lccccccc}
\multicolumn{8}{c}{OLS} \\
 & 1970 & 1975 & 1980 & 1985 & 1990 & 1995 & 2000\\
\hline
NS$^{tot}$ & 0.05 (1.00) & 0.03 (1.00) & 0.04 (1.00) & 0.03 (1.00) & 0.04 (1.00) & 0.03 (1.00) & 0.03 (1.00)\\
ANNS$^{tot}$ & 0.05 (1.00) & 0.05 (0.99) & 0.04 (1.00) & 0.04 (1.00) & 0.03 (1.00) & 0.03 (1.00) & 0.05 (1.00)\\
WCC$^{tot}$ & 0.05 (0.99) & 0.05 (0.99) & 0.04 (1.00) & 0.07 (0.88) & 0.05 (0.99) & 0.05 (0.98) & 0.06 (0.95)\\
\hline
\multicolumn{8}{c}{} \\
\multicolumn{8}{c}{PPML} \\
 & 1970 & 1975 & 1980 & 1985 & 1990 & 1995 & 2000\\
\hline
NS$^{tot}$ & 0.06 (0.96) & 0.10 (0.44) & 0.11 (0.31) & 0.11 (0.27) & 0.08 (0.78) & 0.08 (0.64) & 0.08 (0.72)\\
ANNS$^{tot}$ & 0.98 (0.00) & 0.99 (0.00) & 0.99 (0.00) & 0.97 (0.00) & 0.94 (0.00) & 0.94 (0.00) & 0.94 (0.00)\\
WCC$^{tot}$ & 0.45 (0.00) & 0.36 (0.00) & 0.32 (0.00) & 0.38 (0.00) & 0.32 (0.00) & 0.29 (0.00) & 0.34 (0.00)\\
\hline
\multicolumn{8}{c}{} \\
\multicolumn{8}{c}{ZIP} \\
 & 1970 & 1975 & 1980 & 1985 & 1990 & 1995 & 2000\\
\hline
NS$^{tot}$ & 0.06 (0.96) & 0.07 (0.92) & 0.08 (0.77) & 0.09 (0.6) & 0.06 (0.94) & 0.08 (0.73) & 0.05 (0.98)\\
ANNS$^{tot}$ & 0.95 (0.00) & 0.93 (0.00) & 0.94 (0.00) & 0.87 (0.00) & 0.83 (0.00) & 0.85 (0.00) & 0.83 (0.00)\\
WCC$^{tot}$ & 0.48 (0.00) & 0.41 (0.00) & 0.38 (0.00) & 0.40 (0.00) & 0.33 (0.00) & 0.32 (0.00) & 0.36 (0.00)\\
\hline
\end{tabular}
\caption{Two-sample Kolmogorov-Smirnov test statistics. Null hypothesis: predicted and observed node-statistics sequences come from the same distribution. P-values in parentheses.}\label{table:ks_tests}
\end{footnotesize}
\end{table}


\newpage


\vspace{-0.5cm}

\vspace{-0.5cm}

\end{document}